\documentclass[twocolumn]{aastex61}

\pdfoutput=1 %for arXiv submission
\usepackage{amsmath,amstext,gensymb}
\usepackage[T1]{fontenc}
\usepackage{apjfonts,color}
\usepackage[figure,figure*]{hypcap}

\newcommand{\teff}{$T_{\mathrm{eff}}$}

\newcommand{\numax}{$\nu_{\mathrm{max}}$}
\newcommand{\dnu}{$\Delta\nu$}

\newcommand{\rsol}{R$_\odot$}

\newcommand{\kepler}{\textit{Kepler}}

\def\be{\begin{eqnarray}}   \def\ee{\end{eqnarray}}
\shorttitle{TESS-HERMES DR1}
\shortauthors{Sharma et al.}
\begin{document}
\title{The TESS-HERMES survey Data Release 1: high-resolution spectroscopy of the TESS southern continuous viewing zone}
\author{Sanjib Sharma}
\affiliation{Sydney Institute for Astronomy, School of Physics,
University of Sydney, NSW 2006, Australia}
\author{Dennis Stello}
\affiliation{School of Physics, University of New South Wales, Sydney, NSW 2052, Australia}
\affiliation{Stellar Astrophysics Centre, Department of Physics and Astronomy, Aarhus University, DK-8000 Aarhus C, Denmark}
\affiliation{Sydney Institute for Astronomy, School of Physics,
University of Sydney, NSW 2006, Australia}
\author{Sven Buder}
\affiliation{Max Planck Institute  for Astronomy (MPIA), Koenigstuhl 17, D-69117 Heidelberg}
\affiliation{Fellow of the International Max Planck Research School for Astronomy \& Cosmic Physics at the University of Heidelberg}
\author{Janez Kos}
\affiliation{Sydney Institute for Astronomy, School of Physics,
University of Sydney, NSW 2006, Australia}
\author{Joss Bland-Hawthorn}
\affiliation{Sydney Institute for Astronomy, School of Physics, University of Sydney, NSW 2006, Australia}
\author{Martin Asplund}
\affiliation{Research School of Astronomy \& Astrophysics, Australian National University, ACT 2611, Australia}
\author{Ly Duong}
\affiliation{Research School of Astronomy \& Astrophysics, Australian National University, ACT 2611, Australia}
\author{Jane Lin}
\affiliation{Research School of Astronomy \& Astrophysics, Australian National University, ACT 2611, Australia}
\author{Karin Lind}
\affiliation{Max Planck Institute  for Astronomy (MPIA), Koenigstuhl 17, D-69117 Heidelberg}
\affiliation{Department of Physics and Astronomy, Uppsala University, Box 516, SE-751 20 Uppsala, Sweden}
\author{Melissa Ness}
\affiliation{Max Planck Institute  for Astronomy (MPIA), Koenigstuhl 17, D-69117 Heidelberg}
\author{Daniel Huber}
\affiliation{Institute for Astronomy, University of Hawai`i, 2680 Woodlawn Drive, Honolulu, HI 96822, USA}
\affiliation{Sydney Institute for Astronomy (SIfA), School of Physics, University of Sydney, NSW 2006, Australia}
\affiliation{SETI Institute, 189 Bernardo Avenue, Mountain View, CA 94043, USA}
\affiliation{Stellar Astrophysics Centre, Department of Physics and Astronomy, Aarhus University, Ny Munkegade 120, DK-8000 Aarhus C, Denmark}
\author{Tomaz Zwitter}
\affiliation{Faculty of Mathematics and Physics, University of Ljubljana, Jadranska 19, 1000 Ljubljana, Slovenia}
\author{Gregor Traven}
\affiliation{Faculty of Mathematics and Physics, University of Ljubljana, Jadranska 19, 1000 Ljubljana, Slovenia}
\author{Marc Hon}
\affiliation{School of Physics, University of New South Wales, Sydney, NSW 2052, Australia}
\author{Prajwal R. Kafle}
\affiliation{International Centre for Radio Astronomy Research (ICRAR), The University of Western Australia, 35 Stirling Highway, \\Crawley, WA 6009, Australia}
\author{Shourya Khanna}
\affiliation{Sydney Institute for Astronomy, School of Physics,
University of Sydney, NSW 2006, Australia}
\author{Hafiz Saddon}
\affiliation{School of Physics, University of New South Wales, Sydney, NSW 2052, Australia}
\author{Borja Anguiano}
\affiliation{Department of Astronomy, University of Virginia, Charlottesville, VA 22904-4325, USA}
\affiliation{Department of Physics \& Astronomy, Macquarie University, Sydney, NSW 2109, Australia}
\author{Andrew R. Casey}
\affiliation{School of Physics \& Astronomy, Monash University, Clayton 3800, Victoria, Australia}
\affiliation{Faculty of Information Technology, Monash University, Clayton 3800, Victoria, Australia}
\author{Ken Freeman}
\affiliation{Research School of Astronomy \& Astrophysics, Australian National University, ACT 2611, Australia}
\author{Sarah Martell}
\affiliation{School of Physics, University of New South Wales, Sydney, NSW 2052, Australia}
\author{Gayandhi M. De Silva}
\affiliation{Sydney Institute for Astronomy, School of Physics, University of Sydney, NSW 2006, Australia}
\affiliation{Australian Astronomical Observatory, 105 Delhi Rd, North Ryde, NSW 2113, Australia}
\author{Jeffrey D. Simpson}
\affiliation{Australian Astronomical Observatory, 105 Delhi Rd, North Ryde, NSW 2113, Australia}
\author{Rob A. Wittenmyer}
\affiliation{University of Southern Queensland, Computational
Engineering and Science Research Centre, Toowoomba, Queensland 4350,
Australia}
\author{Daniel B. Zucker}
\affiliation{Department of Physics \& Astronomy, Macquarie University, Sydney, NSW 2109, Australia}
\affiliation{Research Centre in Astronomy, Astrophysics \& Astrophotonics, Macquarie University, Sydney, NSW 2109, Australia}
\affiliation{Australian Astronomical Observatory, 105 Delhi Rd, North Ryde, NSW 2113, Australia}

\begin{abstract}
The Transiting Exoplanet Survey Satellite (TESS) will provide high precision time-series photometry for millions of stars with at least a half-hour cadence. Of particular interest are the circular regions of 12-degree radius centered around the ecliptic poles that will be observed continuously for a full year. Spectroscopic stellar parameters are desirable to characterize and select suitable targets for TESS, whether they are focused on exploring exoplanets, stellar astrophysics, or Galactic archaeology.  Here, we present spectroscopic stellar parameters ($T_{\rm eff}$, $\log g$, [Fe/H], $v \sin i$, $v_{\rm micro}$) for about 16,000 dwarf and subgiant stars in TESS' southern continuous viewing zone.  For almost all the stars, we also present Bayesian estimates of stellar properties including distance, extinction, mass, radius, and age using theoretical isochrones.  Stellar surface gravity and radius are made available for an additional set of roughly 8,500 red giants.  All our target stars are in the range $10<V<13.1$. Among them, we identify and list 227 stars belonging to the Large Magellanic Cloud.  The data were taken using the the High Efficiency and Resolution Multi-Element Spectrograph (HERMES, R $\sim 28,000$) at the Anglo-Australian Telescope as part of the TESS-HERMES survey. Comparing our results with the TESS Input Catalog (TIC) shows that the TIC is generally efficient in separating dwarfs and giants, but it has flagged more than hundred cool dwarfs ($T_{\rm eff}< 4800$ K) as giants, which ought to be high-priority targets for the exoplanet search.  The catalog can be accessed via \href{http://www.physics.usyd.edu.au/tess-hermes/}{http://www.physics.usyd.edu.au/tess-hermes/}, or at \href{https://archive.stsci.edu/prepds/tess-hermes/}{MAST}.
\end{abstract}
\keywords{catalogues --- surveys --- stars: fundamental
parameters --- planetary systems  --- Galaxy: stellar
content --- techniques: spectroscopic}

\section{Introduction}
The precise photometric time-series observations by \kepler\ have made it one of NASA's most successful missions.  Its successor, the Transiting Exoplanet Survey Satellite (TESS), is expected to be launched in March 2018.  TESS will obtain high-precision time-resolved photometric data for millions of stars covering most of the sky.  Whilst TESS' primary role is to detect planets around nearby stars, its data will likely provide a huge boost to a variety of other lines of research, such as asteroseismology and stellar rotation; just like \kepler\ transformed these fields by observing about 150,000 stars in a roughly 100 square degree patch of sky for four years.  Unlike \kepler, the TESS observing strategy will result in regions of sky being observed between 27.4 days to 356.2 days, the longest in the so-called Continuing Viewing Zones (CVZs) around the ecliptic poles, which each cover approximately 450 square degrees, or more than four times the size of the \kepler\ field \citep{2014SPIE.9143E..20R}.  TESS' entire field of view will be observed at 30-minute cadence and at least 200,000 selected stars will be observed in a 2-minute high-cadence mode.  The long duration of observation of the CVZs makes them particularly valuable for many aspects of time-series-based astrophysics research.  As a result, the CVZs are expected to contain the highest density of high-priority, as well as high-cadence, targets to follow with ancillary observations from ground and space.

In this paper we present the first results of the spectroscopic TESS-HERMES survey on the 4-m Anglo-Australian telescope (AAT).  The goal of the survey is to obtain a high-resolution spectrum for all bright stars in TESS' southern CVZ -- the first of the two CVZs to be observed by TESS.  By focusing on the southern CVZ, the TESS-HERMES survey aims to facilitate and enhance the use of the TESS data for the stars with the longest, hence best, TESS observations.  Our results can further serve for calibration of other spectroscopic surveys, such as the low-resolution FunnelWeb\footnote{https://funnel-web.wikispaces.com/The+FunnelWeb+Survey}, which will observe the entire southern hemisphere.  The purpose of this paper and associated data release in particular, is to support the selection of TESS targets prior to launch, in time for the first NASA call for TESS Guest Investigator proposals, for which the second data release of Gaia \citep{2016A&A...595A...1G} expected in April 2012, will be too late.

In Section~\ref{program} we describe the TESS-HERMES survey including its target selection strategy, the data reduction and quality, and the spectroscopic analysis.  In Section~\ref{properties} we present the method used to derive the fundamental stellar properties inferred from the spectroscopic results (\teff, $\log g$, and [Fe/H]) using grid-based modeling.  We explain the content of the TESS-HERMES catalog associated with this paper in Section~\ref{catalog}, which includes a discussion of the separation of dwarfs and subgiants.  We finally conclude with a summary in Section~\ref{summary}.

\section{The TESS-HERMES survey}\label{program}

\subsection{Observations, target selection, and completeness}\label{sec:observations}

The observations were obtained with the 4-m AAT
located at Siding Spring Observatory in Australia. We use the High Efficiency and Resolution Multi-Element Spectrograph (HERMES), which can obtain spectra of up to 360 science targets simultaneously \citep{2015JATIS...1c5002S,2012SPIE.8446E..0WH,2011SPIE.8125E..04B,2010SPIE.7735E..09B}.

We use 2MASS \citep{2006AJ....131.1163S} as our base input catalog. For positions and proper motions we use the values from UCAC4 \citep{2013AJ....145...44Z}, which is cross matched to 2MASS.
We choose stars only with good quality photometry (Table \ref{tab:tmassqual}), and use the following formula to convert 2MASS infrared magnitudes to a magnitude in $V$ band (denoted $V_{JK}$), which is used by the GALAH survey \citep{2017MNRAS.465.3203M}.
\begin{equation}
V_{JK}=Ks + 2.0 (J-Ks+0.14)+0.382\exp[(J-Ks-0.2)/0.50].
\label{equ:vjk}
\end{equation}
\begin{figure}
\centering \includegraphics[width=0.48\textwidth]{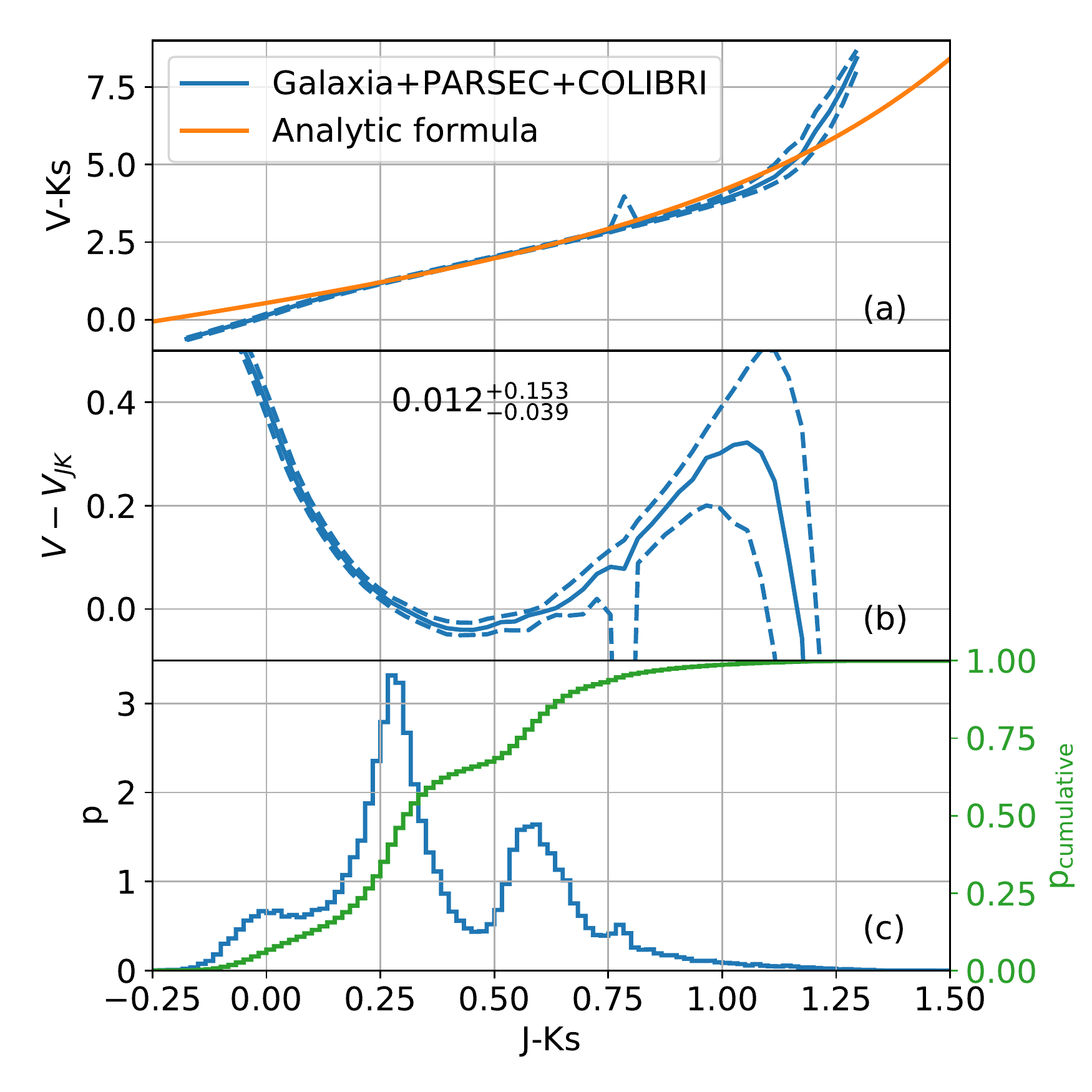}\caption{Comparison of $V$ magnitude in the Johnson Cousins system \citep{1990PASP..102.1181B} with that of $V_{JK}$ estimated from 2MASS $J$ and $Ks$ using \autoref{equ:vjk}. The comparison is done with a stellar population simulated by {\sl Galaxia} ($|b|>10$, $V<14$) using PARSEC-COLIBRI stellar isochrones and unredddened colors.
(a) $V-Ks$ color as function of $J-Ks$ color of simulated population (blue curves) and as predicted by \autoref{equ:vjk}. The solid blue curves show the median, while the dashed curves show the 16th and 84th percentiles.
(b) $V-V_{JK}$ as function of $J-Ks$ color using the same annotation as in panel (a). Its median and the scatter is also listed on the panel.
(c) The color distribution of the simulated stellar population, both differential (blue curve) and cumulative (green curve).
\label{fig:vjk}}
\end{figure}
We tested the accuracy of the formula using a synthetic stellar population simulated by {\sl Galaxia} \citep{2011ApJ...730....3S} with PARSEC-COLIBRI stellar isochrones \citep{2017ApJ...835...77M}. \autoref{fig:vjk}a shows that the formula provides a good match to the $V-Ks$ color for a wide range of $J-Ks$ colors all the way from hot dwarfs to cool giants.
In the range $0.22<J-Ks<0.7$ (corresponds to approximately $4500<T_{\rm eff}/{\rm K}<6500$), which encloses 67\% of simulated stars with $V<14$ and $|b|>10$, the formula is accurate to better than 0.05 mag (\autoref{fig:vjk}bc). Outside this range, the formula  becomes increasingly inaccurate with $V$ magnitude being systematically underestimated for $J-Ks<0.22$ and $0.7< J-Ks<1.15$.
The large uncertainty at $(J-Ks) \sim 0.78$ is
due to cool K and M dwarfs, for which the color does not
vary with temperature.

\begin{table}
\caption{2MASS quality selection criteria}
\begin{tabular}{@{}|l|l|l|l|}
\hline
Flag  & Criterion & Description\\
\hline
Qflag & $=$ 'AAA'    & photometric quality in $J,H,K$ better than B\\
Bflag & $=$ '111'    & blend flag \\
Cflag & $=$ '000'    & contamination flag\\
Xflag & $=$ 0        & extended source contamination flag\\
Aflag & $=$ 0        & known solar system object flag \\
prox  & $>$ 6 arcsec & distance to nearest star\\
\hline
\end{tabular}
\label{tab:tmassqual}
\end{table}

Our main targets are bright stars in order to support both asteroseismic and planet host targets. Due to `cross talk' in the spectrograph, one can only observe stars spanning about three magnitudes in brightness within a single exposure.  To optimize the fibre allocation towards scientifically interesting targets for asteroseismology and exoplanet research, we adopted a target magnitude range of $10<V<13.1$.  On average this provides 350 targets per observing field, which is two degrees in diameter.

The exposure time was chosen to achieve a S/N of 100 per resolution element (or 50 per pixel). The Galah survey, which also uses HERMES, has shown that $\mathrm{S/N} \sim 100$ can be achieved in 1-hour exposures for stars with $V \sim 14$ mag \citep{2015MNRAS.449.2604D, 2015JATIS...1c5002S, 2017MNRAS.465.3203M}.
When we are not background limited, we can achieve $\mathrm{S/N}=107$ for stars as faint as $V=13.1$ mag with exposure times of $T_{\rm exp} =30$ min. Assuming 18 min of overhead for calibration frames, readout, and plate change, this leads to a 48 min duty cycle per field.

\begin{figure*}
\centering \includegraphics[width=0.98\textwidth]{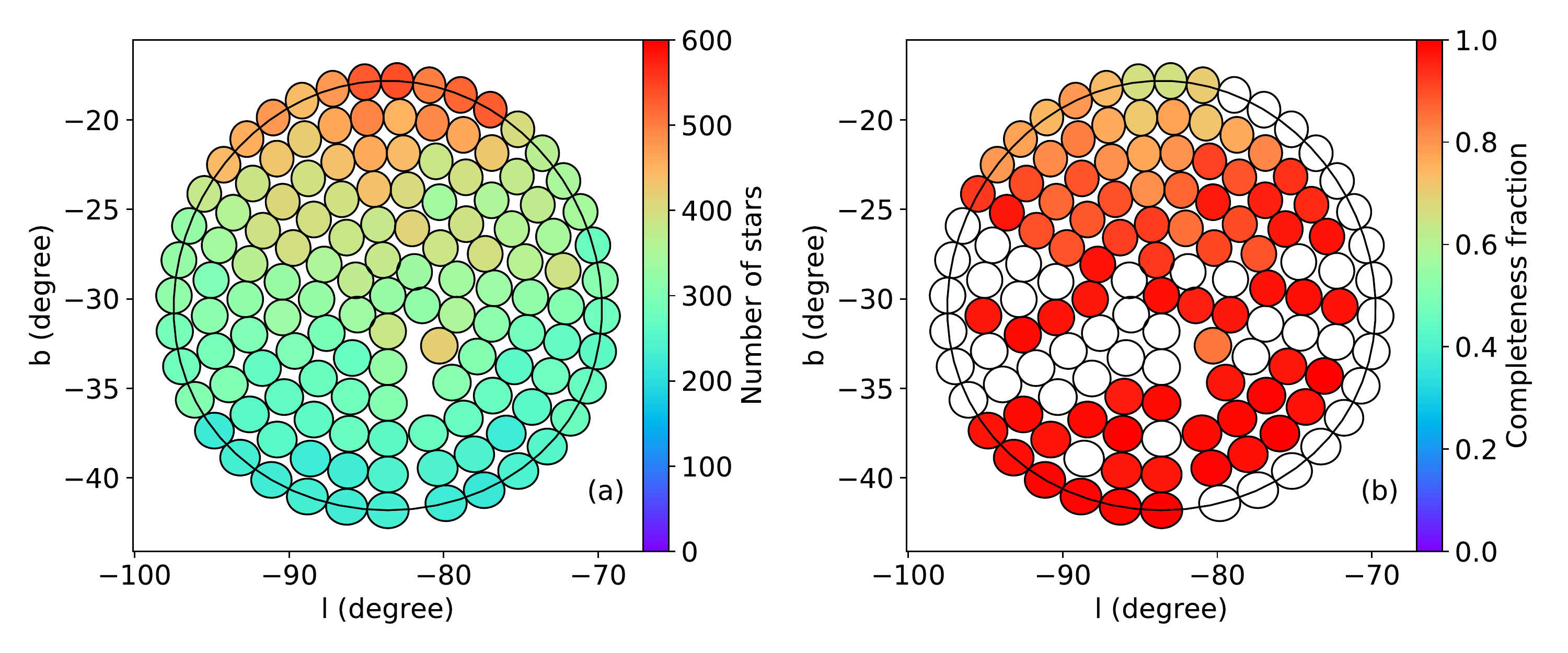}\caption{Tiling of the southern TESS CVZ using 1-degree radius fields (Galactic coordinates).
The large solid black circle indicates the 12-degree radius region that will be observed for 356.2 days (known as the CVZ).  The fields are non-overlapping and lie all within 13-degrees from the ecliptic pole. (a) Color-code shows the number of stars with $10<V<13.1$ in each field.  (b) Color-code shows the completeness fraction of observed stars (with $10<V<13.1$). White circles denotes unobserved fields.
\label{fig:tess_fields}}
\end{figure*}

The total area of the CVZ is 452 square degrees. We tiled it with non-overlapping circles (2-degree diameter), which offers minimum redundancy (Figure~\ref{fig:tess_fields}a).
Our tiling scheme fitted 130 fields centered within a 12-degree radius around the south ecliptic pole out of which 93 are within a 11 degree radius covering 292 square degrees. This leads to a CVZ sky coverage fraction of 0.77.

The CVZ is best observed from the AAT around 15th of December.
We were allocated 17 nights in the period 13th Dec 2016 to 16th Jan 2017. About 50\% of the time was lost due to bad weather. In the remaining time we collected data for 81 fields (\autoref{fig:tess_fields}b), all of which are presented in the current paper. This comprises 25,723 stars in total, although currently our spectral analysis pipeline has delivered stellar parameters for only 24,741 of those.

\begin{figure}
\centering \includegraphics[width=0.46\textwidth]{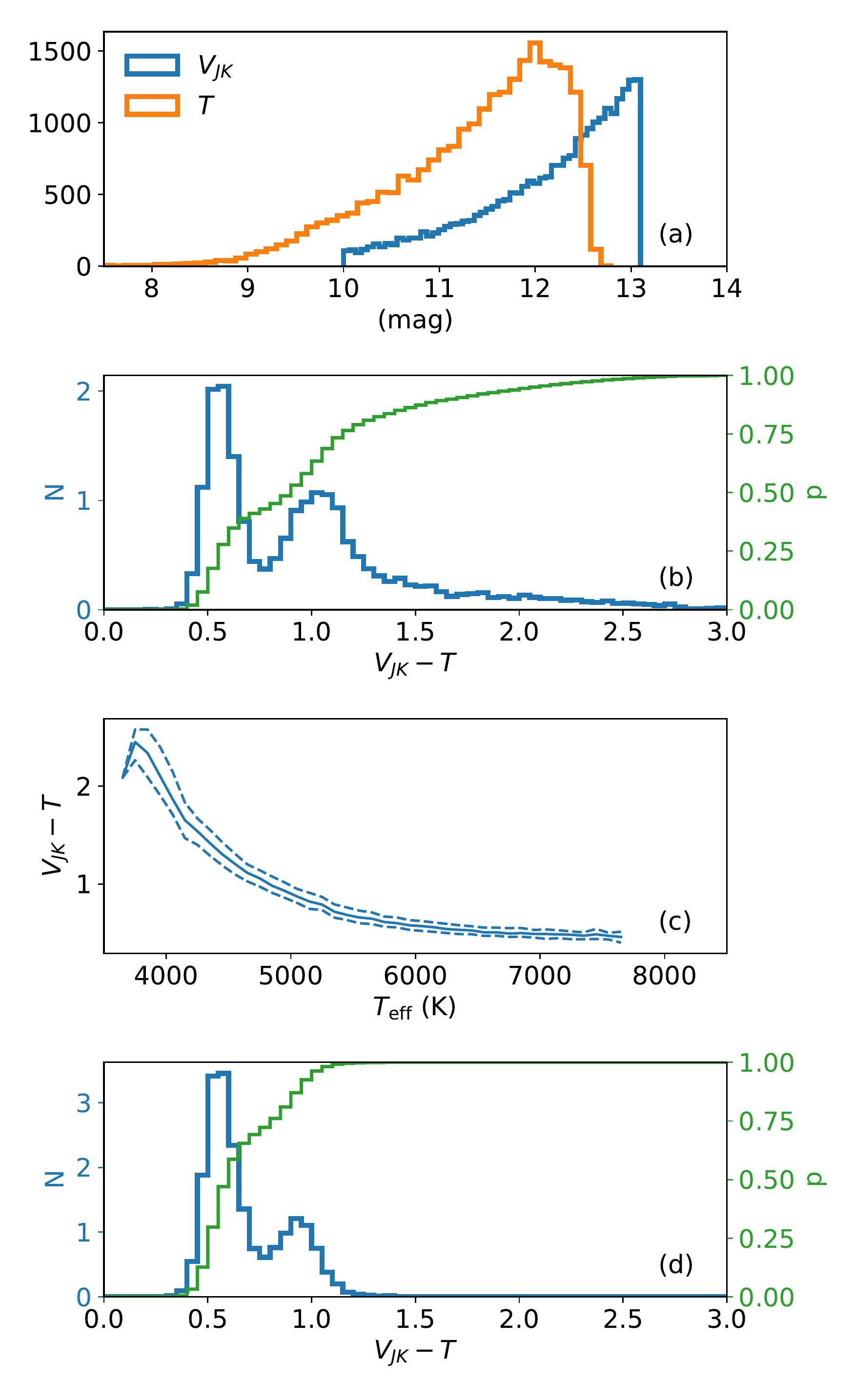}\caption{ (a) $V_{JK}$ and $T$ distributions of our full sample.  (b) Distribution of $V_{JK}-T$ for stars with $T<11$ and $10 < V_{JK} < 13.1$ (blue curve).  The associated cumulative distribution is shown by the green curve.
(c) Relation between $V_{JK}$ and \teff\ for our sample. The solid curve shows the 50th percentile of the distribution, while the dashed curves show the 16th and 84th percentiles.
(d) Like (b) but restricted to hot stars ($T_{\rm eff}> 4800\,$K).
\label{fig:vmag_tmag}}
\end{figure}

Our target selection is made in terms of $V_{JK}$ magnitude (\autoref{equ:vjk}) with a magnitude cut at 13.1 (\autoref{fig:vmag_tmag}a, blue curve). However, the primary photometric band for TESS is $T$, which makes it important to understand the completeness of our survey in terms of the $T$ band. This is particularly important for bright dwarfs with $T<12$, which are the prime targets of the TESS mission.
We obtain $T$ for our targets from the TESS Input Catalog (TIC)\footnote{In this paper we refer to TIC-5.} \citep{2017arXiv170600495S} by cross matching 2MASS IDs.
In \autoref{fig:vmag_tmag}a (orange curve) we see clearly that the $T$ magnitude distribution of our sample falls off beyond 12th mag, meaning the sample is not magnitude complete (in $T$) for fainter stars.  To study the dependence on completeness of temperature (color), we analyze the distribution of the $V_{JK}-T$ color for the target stars, and to ensure completeness in the $T$ band, we look only at stars with $T<11$ mag.
In \autoref{fig:vmag_tmag}b the 50th percentile along the cumulative distribution (thin green curve) shows that the median color is $V_{JK}-T \simeq 1$, suggesting that our $V_{JK}<13.1$ magnitude limited sample is only complete till $T<12.1$ for stars with $V_{JK}-T < 1$. This also shows that we are missing all stars with $V_{JK}-T > 1$.
To convert this color-threshold of the completeness to an effective temperature, we show in \autoref{fig:vmag_tmag}c the relation between $V_{JK}- T$ and \teff. The relation between $V_{JK}- T$ and \teff\ for our sample is fairly tight as illustrated by the narrow band between the two dashed lines, which encapsulate 64\% of the stars.
We see that the $V_{JK}-T = 1$ color corresponds roughly to \teff\ $=4800\,$K.
So now, if we restrict the sample to be stars hotter than $T_{\rm eff}> 4800 K$, we see from \autoref{fig:vmag_tmag}d (thin green curve) that our sample is 98\% complete for stars below the same color threshold of $V_{JK}- T$ < 1, corresponding to $T<12.1$.

\subsection{Data reduction and quality}\label{sec:reduction}

The TESS-HERMES survey uses the same instrument as the Galah survey \citep{2015MNRAS.449.2604D,2017MNRAS.465.3203M} and follow a similar observing strategy. Hence, we use the same reduction pipeline as Galah to perform the data reduction from the raw CCD images to the final calibrated spectra. For this, a custom IRAF-based pipeline has been made that performs initial quality checks, optimal extraction, reduction, and basic analysis of spectra. It is described in detail in \cite{2017MNRAS.464.1259K}. Only minor modifications of the pipeline have been made since, mostly aimed to improve (1) sky subtraction, using a higher order model for the sky brightness variation across the 2$^\circ$ field, (2) telluric absorption fitting by implementing a better $\chi^2$ minimization algorithm, and (3) the precision of the wavelength solution. Wavelength solutions now adapt better to regions in the spectrograph's red and green bands with few arc lines where the solution previously was not perfect.

For every observed star the pipeline produces a one-dimensional normalized spectrum in each of four bands (blue 4718-4903~\AA; green 5649-5873~\AA; red 6481-6739~\AA; IR 7590-7890~\AA) together with an estimate of the spectral uncertainty at each wavelength, and four basic parameters ($T_{\mathrm{eff}}$, $\log g$, $\mathrm{[Fe/H]}$, and radial velocity).
The basic stellar parameters, the radial velocity, and continuum normalization are calculated with a custom made pipeline, named GUESS, by matching the observed normalized spectra to synthetic templates . A grid of AMBRE synthetic spectra is used for this purpose \citep{2012A&A...544A.126D}.
The reduction pipeline also combines all consecutive exposures of a field as well as observations from different epochs, if they exist. The combined spectra are used in the following spectroscopic analysis (Section \ref{spec_anal}).

\begin{figure}
\centering \includegraphics[width=0.48\textwidth]{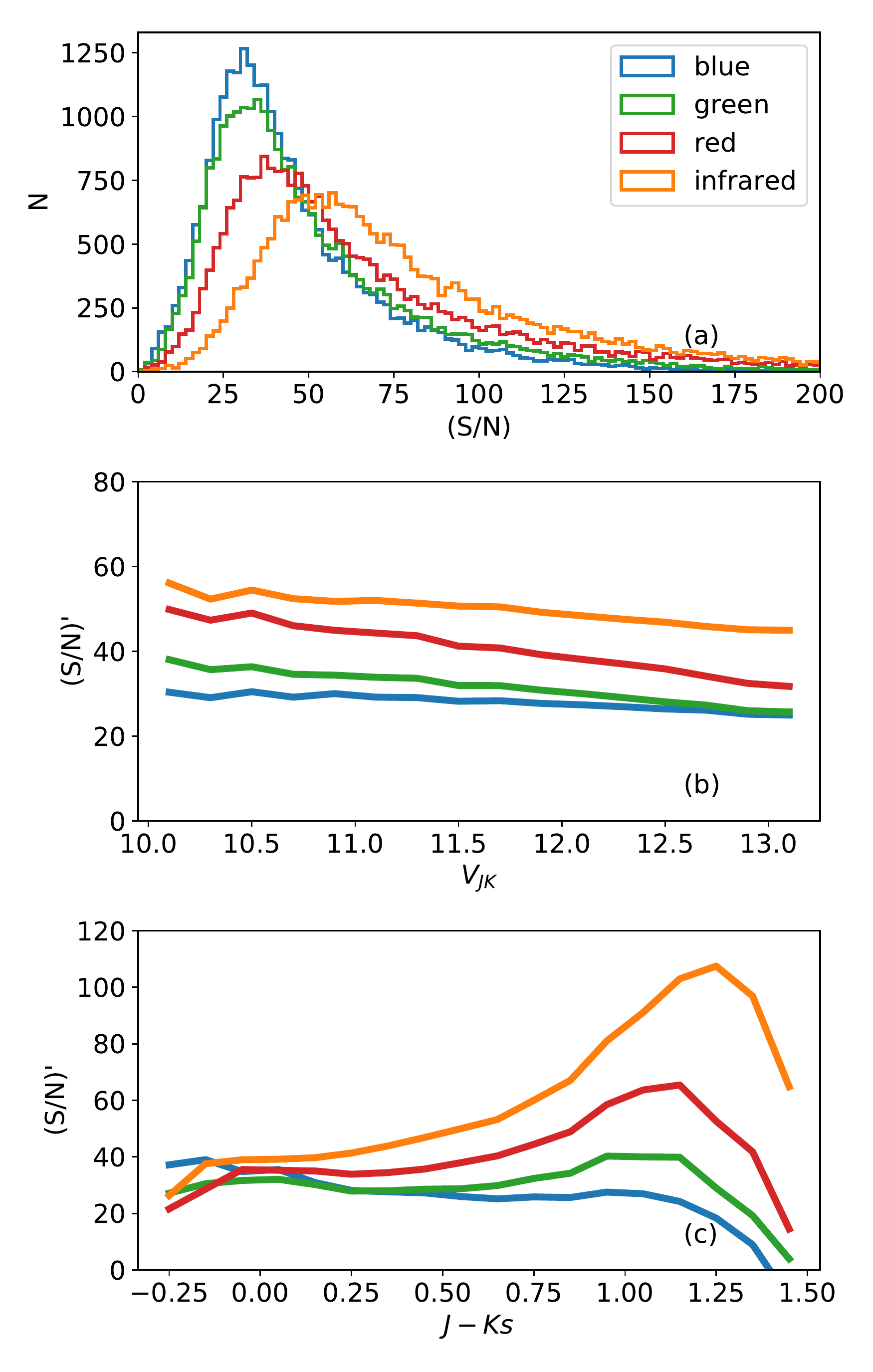}\caption{Signal to noise per pixel (S/N) for the different HERMES wavelength bands (S/N per resolution element is about twice as high). (a) Sample distribution of S/N. (b) Mean S/N' (scaled signal to noise, which accounts for changes in exposure time, magnitude, and airmass, see \autoref{equ:snoise}) as function of $V_{JK}$ magnitude.  (c) Mean S/N' as function of $J-Ks$ color.  The green channel is close to $V$ band and hence shows the least dependence on color. \label{fig:snr_iraf}}
\end{figure}

In \autoref{fig:snr_iraf} we show the quality
of our reduced spectra. The distribution of signal-to-noise per pixel (S/N) for the four different wavelength bands (channels) is shown in \autoref{fig:snr_iraf}a.
The S/N depends on magnitude $V_{JK}$, exposure time $T_{\rm exp}$, and air mass $\sec z$ ($z$ being the zenith angle). We therefore derive a normalized signal to noise
\be
{\rm (S/N)'}=\frac{\rm (S/N)}{\sqrt{(T_{\rm exp}/1 {\rm\,hr}) 10^{0.4(14-V+(\sec z-1))}}},
\label{equ:snoise}
\ee
which should take out these dependence (\autoref{fig:snr_iraf}b).
For this, we assumed that Poisson noise of the signal is the dominant noise source.  Indeed, (S/N)$'$ is seen to have very little dependence on apparent magnitude $V_{JK}$ (\autoref{fig:snr_iraf}b). In particular, the smallest dependence on color $J-Ks$ is shown by the green channel because this is the one most similar to the $V_{JK}$ band (\autoref{fig:snr_iraf}c).  For this reason, we adopt the S/N in the green channel as our default signal-to-noise and use it to characterize the spectroscopic properties of our sample.

\subsection{Spectroscopic analysis}\label{spec_anal}

The spectroscopic stellar parameters have been estimated with a combination of classical spectrum synthesis for a representative reference set of stars and a data-driven approach to propagate the high-fidelity parameter information with higher precision onto all the stars in the TESS-HERMES survey. This approach was chosen as part of the joint spectral analysis with the Galah survey and the K2-HERMES survey (Sharma et al. in prep.), as briefly mentioned in \cite{2017MNRAS.465.3203M}.

\subsubsection{Building a reference set}

In the first step, we create a reference set of stars, which we use for both external validation and as training set for the parameter propagation with \textit{The Cannon} \citep{2015ApJ...808...16N} in \autoref{sec:cannon}. A combination of different samples of stars with external estimates of parameters were chosen, including Gaia FGK benchmark stars with their (fundamental) parameters \cite{2015A&A...582A..49H,2014A&A...564A.133J}, member stars of open and globular clusters with intrinsically very similar chemical composition, as well as stars with asteroseismic information from the K2 campaign C1 \citep{2017ApJ...835...83S}.

Next, we use the spectrum synthesis code Spectroscopy Made Easy (SME) by \cite{2017A&A...597A..16P} to analyse the reference set.  For this, we carefully selected spectral segments containing the Balmer lines as well as blend-free Fe/Sc/Ti neutral and ionised lines for which the wavelengths, excitation energies, and oscillator strengths are confirmed by laboratory experiments. Starting from a set of initial stellar parameters delivered by the reduction pipeline (\autoref{sec:reduction}) the parameters $T_{\rm eff}$, $\log g$, ${\rm [Fe/H]}$, $v_\text{mic}$, $v_\text{mac+rot}$, and $v_\text{rad}$ were optimized with a Levenberg-Marquardt algorithm based on the $\chi^2$ of the observations and on-the-fly synthesized spectra. Syntheses were performed using 1D MARCS model atmospheres \citep{2008A&A...486..951G} with non-LTE corrections by \cite{2012MNRAS.427...50L} for iron, and with standard composition, which was solar scaled based on the values by \cite{2007SSRv..130..105G} with a gradual alpha-enhancement up to $0.4\,\mathrm{dex}$.
Wherever the parameter space for $T_{\rm eff}$, $\log g$, and ${\rm [Fe/H]}$ was not well covered by our reference set, stars with high S/N spectra and reliable parameters were chosen to fill the gap, if available. Because of the strong influence of molecular absorption bands and the limitation to 1D atmospheres for cool stars, as well as decreasing line strength for hot stars, we only include dwarfs here with $4900 < T_{\rm eff}/\mathrm{K} < 7500$ but will address the remaining stars in later releases. The final sample for the reference set consists of 2191 stars covering dwarfs, turn-off stars, and giants.

\begin{figure}
\centering
\includegraphics[width=0.45\textwidth]{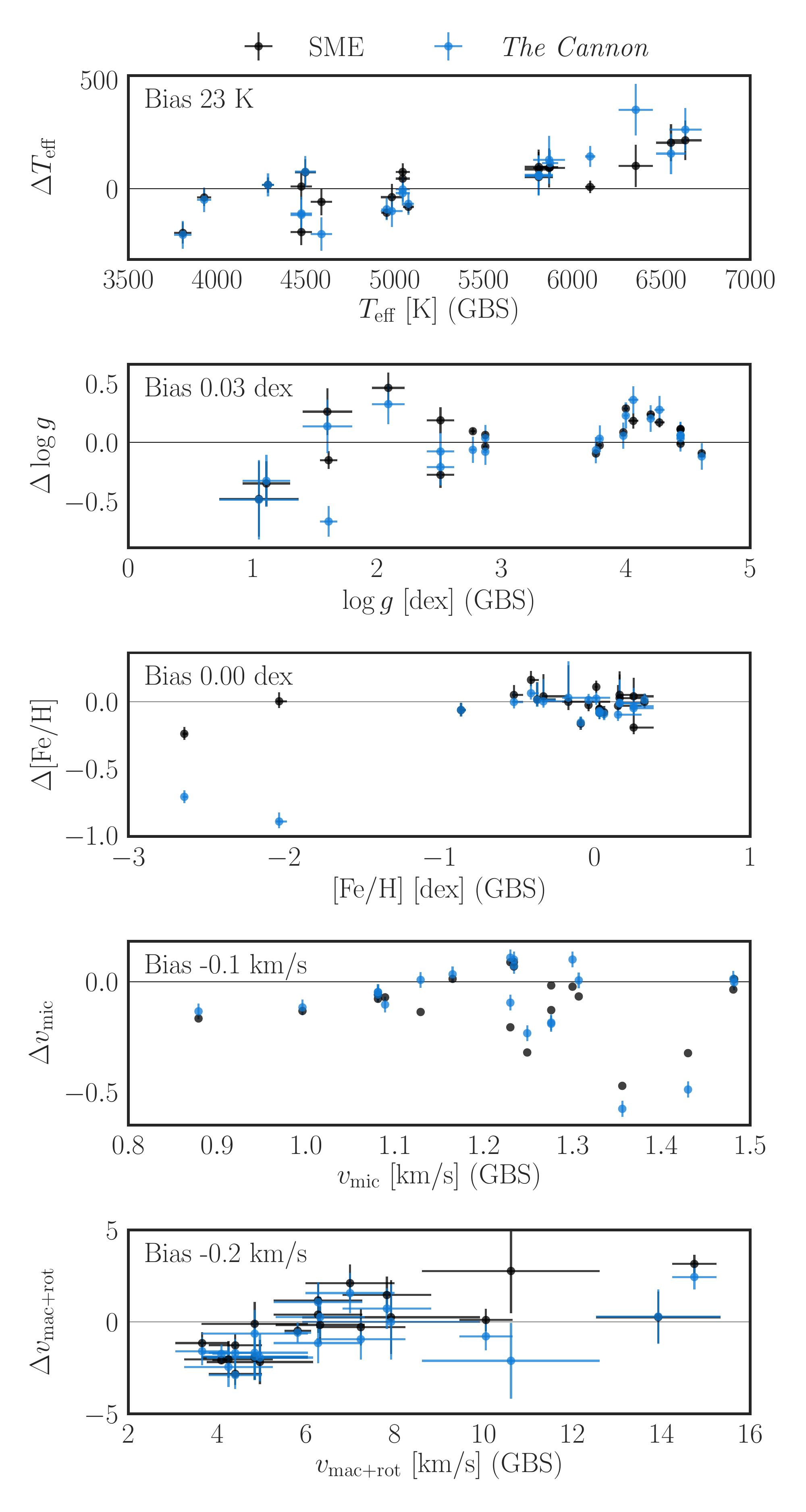}\caption{Differences between (fundamental) parameters of the Gaia FGK benchmark stars (GBS) \citep{2015A&A...582A..49H,2014A&A...564A.133J} and our HERMES-based results (plotted as 'GBS' -- 'HERMES')
for $T_{\rm eff}$, $\log g$, ${\rm [Fe/H]}$, $v_\text{mic}$, and $v_\text{mac+rot}$.  Shown are parameters from SME (black) and the reproduced parameters of \textit{The Cannon} (blue). Bias correction have been applied to all $\log g$ and [Fe/H] values (see text). The biases shown are median values of the difference, and the values are: $T_{\rm eff}$ ($23\pm112$ K), $\log g$ ($0.03\pm0.22$ dex), ${\rm [Fe/H]}$ ($0.00\pm0.10$ dex), $v_\text{mic}$ ($-0.1 \pm0.17$km/s), and $v_\text{mac+rot}$ ($-0.2\pm1.7$ km/s). The GBS $v_\text{mac+rot}$ stated here was estimated as the root of the square sum of $v_\text{mac}$ and $v_{\sin i}$ to be comparable with the definition of this study.}\label{fig:GBS_comparison}
\end{figure}

The comparison with fundamental stellar parameters of 21 Gaia FGK benchmark stars, with a focus on the Sun to correct for deviations from the solar composition, revealed biases in both metallicity ($0.12\,\mathrm{dex}$) and surface gravity ($0.15\,\mathrm{dex}$) (benchmarks being higher). The offset in surface gravity was confirmed by comparisons with values derived from asteroseismic measurements. We corrected the biases in [Fe/H] and $\log g$ for the training set before the subsequent \textit{Cannon} analysis (\autoref{sec:cannon}).
We interpret the parameter biases in part as a shortcoming of 1D hydrostatic atmospheres based on comparisons with the STAGGER 3D grid, similar to \cite{2012MNRAS.427...27B}. After correction, the biases of SME-based estimates with respect to the Gaia FGK benchmark stars are negligible, see \autoref{fig:GBS_comparison}.
However, there is considerable scatter and for some parameters we do notice some systematic trends. Our estimated temperatures are slightly hotter for stars with $T_{\rm eff}>6000\,$ K, and slightly cooler for $T_{\rm eff}<5500\,$ K. Our $v_{\rm mac+rot}$ is higher for $v_{\rm mac+rot}<5\,{\rm km/s}$. Furthermore, for ${\rm [Fe/H]}<-1$ \textit{The Cannon} overestimates the metallicity. We note that the two outliers in \teff, [Fe/H], and $v_\textrm{mic}$ in the benchmark comparison are the two very metal-poor stars HD84937 and HD122563. These are both under-represented in the training set, because of a current lack of metal-poor reference stars in the survey volume, for further explanations see \autoref{sec:cannon}.
The rotational velocity, $v \sin i$, is traditionally used to identify fast rotators. However, due to the degeneracy of macro-turbulence and rotational velocity under the assumption of Gaussian line profiles, a combined broadening velocity $v_\text{mac+rot}$ was estimated by SME, and in the catalog we refer to this as $v \sin i$. $v_\text{mac+rot}$ is used to broaden the synthetic spectra together with an assumed Gaussian instrument profile based on resolution maps provided by \cite{2017MNRAS.464.1259K}.
\subsubsection{Extending analysis to all stars: The Cannon}\label{sec:cannon}

\begin{figure*}[ht]
\centering
\begin{minipage}[b]{0.49\textwidth}
\centering
\includegraphics[width=1\textwidth,page=8]{fig6.pdf}\end{minipage}
\begin{minipage}[b]{0.49\textwidth}
\centering
\includegraphics[width=1\textwidth,page=1]{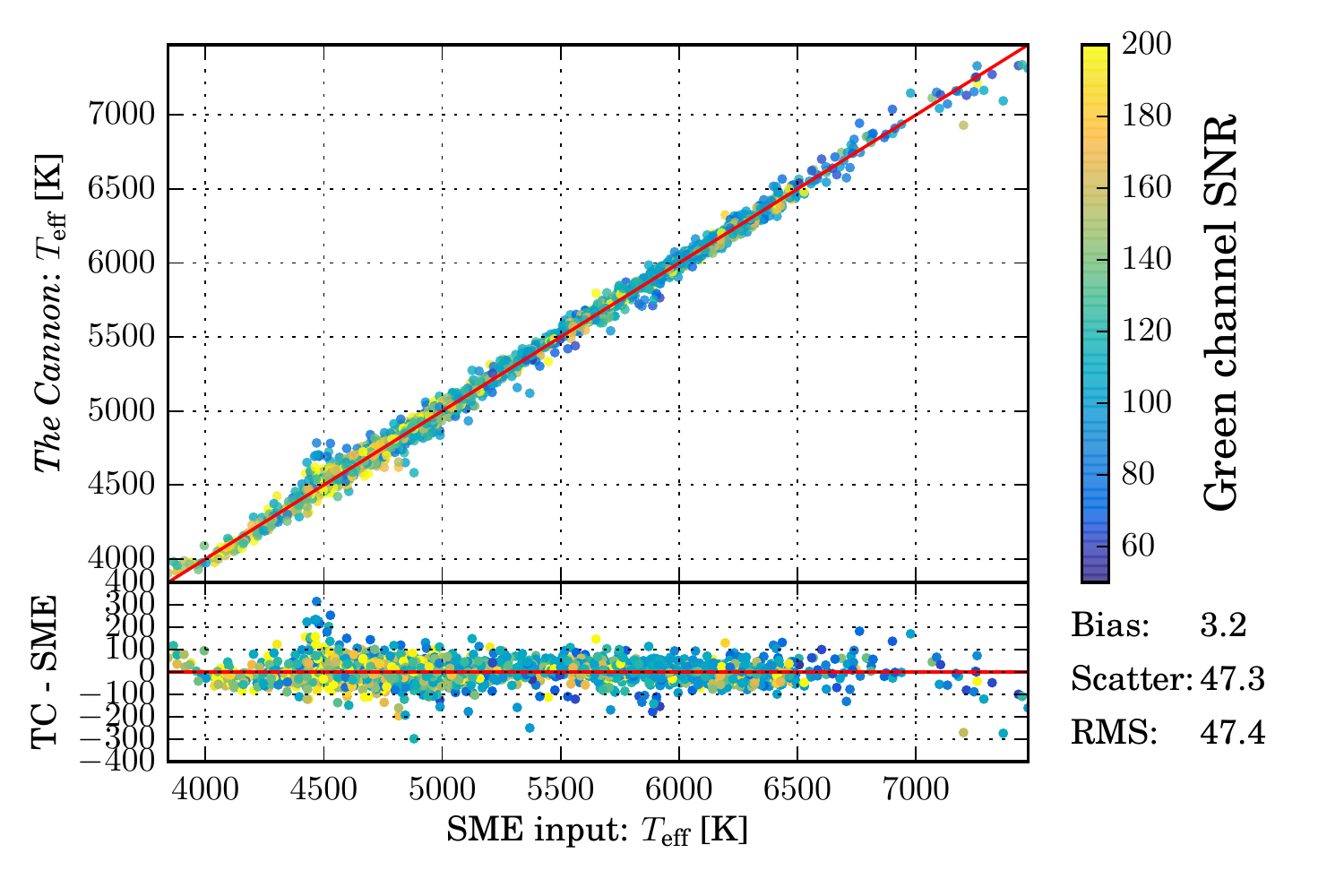}\end{minipage}
\newline
\begin{minipage}[b]{0.49\textwidth}
\centering
\includegraphics[width=1\textwidth,page=2]{fig6.pdf}\end{minipage}
\begin{minipage}[b]{0.49\textwidth}
\centering
\includegraphics[width=1\textwidth,page=3]{fig6.pdf}\end{minipage}
\newline
\begin{minipage}[b]{0.49\textwidth}
\centering
\includegraphics[width=1\textwidth,page=5]{fig6.pdf}\end{minipage}
\begin{minipage}[b]{0.49\textwidth}
\centering
\includegraphics[width=1\textwidth,page=6]{fig6.pdf}\end{minipage}
\caption{Comparison of training set. Top left panel: Comparison of the radial velocity estimated by the reduction pipeline (GUESS), used to shift the TESS-HERMES spectra to the rest wavelength, and the independent analysis by SME for the training set. For better visualization, absolute $v_\text{rad}$ is shown on a logarithmic scale for this panel.
Other 5 panels: Comparison of training set for \textit{The Cannon} for 5 stellar parameters. Shown are the input values from SME and the reproduced values by \textit{The Cannon} as well as their deviation, colored by the S/N per pixel in the green channel.
}\label{fig:leaveouttest}
\end{figure*}

In a second step, we used the data-driven \textit{Cannon} approach to build a quadratic model that connects stellar parameters and the normalized flux for each pixel of a spectrum at the same rest-wavelength grid \citep{2015ApJ...808...16N}.
The normalization and the shift to the rest wavelength is done
by the GUESS pipeline \autoref{sec:reduction}.
Via simultaneous $\chi^2$-optimization for all training set stars, a best model was estimated for each pixel in the so called training step. Thereafter, the TESS-HERMES spectra were analyzed to find the best set of parameters based on this model.
In addition to $(T_{\rm eff}, \log g, {\rm [Fe/H]}, v_{\rm mic}, , v_{\rm mac+rot})$, \textit{The Cannon} also fitted auxiliary labels $A_K$ and [$\alpha$/Fe] for a better overall label estimation, similar to the routine described in \cite{2017MNRAS.465.3203M}.
To validate the performance and reliability of \textit{The Cannon}, we performed leave-20\%-out tests, which generally showed good reproduction of the input (from SME); see  \autoref{fig:GBS_comparison} and \autoref{fig:leaveouttest}.
In  radial velocity comparison, the low number of outliers are confirming the good performance by the reduction pipeline
over a large range of S/N and $v_\text{rad}$.
In the surface gravity comparison,  the influence of strong molecular absorption bands is seen. Because the cool stars share this feature, \textit{The Cannon} is  finding a  compromise in this sparsely sampled parameter space regime, resulting in too high surface gravities for cool giants (similar to too low surface gravities for cool dwarfs).
In the metallicity comparison,  it can be seen that \textit{The Cannon} is overestimating the metallicity of metal-poor stars. Hence, metal-poor stars will be even more metal-poor than estimated by \textit{The Cannon}. However, this affects less than 2\% of the stars with [Fe/H]$<-1.0\,\mathrm{dex}$ of the TESS-HERMES targets.

To describe the reliability of our spectroscopically determined parameters, we use a flagging system as shown in \autoref{tab:spflag}.  Wherever there is not enough training spectra available in a part of the parameter space, \textit{The Cannon} starts to extrapolate. At some stage the extrapolation will not be accurate, hence, in addition to those stars with reduction issues, we set the spflag\_hermes value to 1 for stars that are $3\sigma$ away from the training set.  When the $\chi^2$ value of the best fitting model spectrum is high, we set the flag to 2. Additionally, $\chi^2$ values significantly below the expected values, i.e. with large flux errors in the spectra, are also indicated by a flag 2.  When both the \textit{The Cannon} extrapolates and the $\chi^2$ value is high/low, we set the flag to 3.
\begin{table}
\caption{Spectroscopic quality flag (spflag\_hermes) from the Galah analysis pipeline v1.3}
\begin{tabular}{l l}
\hline\hline
Option & Meaning \\
\hline
0 & Reliable \\
1 & \textit{The Cannon} starts to extrapolate. For some stars\\
& the values could be incorrect. \\
2 & The $\chi^2$ of the best fitting model spectrum is \\
& significantly higher or lower. \\
& Accuracy likely to be compromised. \\
3 & \textit{The Cannon} extrapolates and $\chi^2$ is also higher or lower. \\
& Accuracy likely to be compromised. \\
\hline
\end{tabular}
\label{tab:spflag}
\end{table}

\begin{figure}
\centering \includegraphics[width=0.48\textwidth]{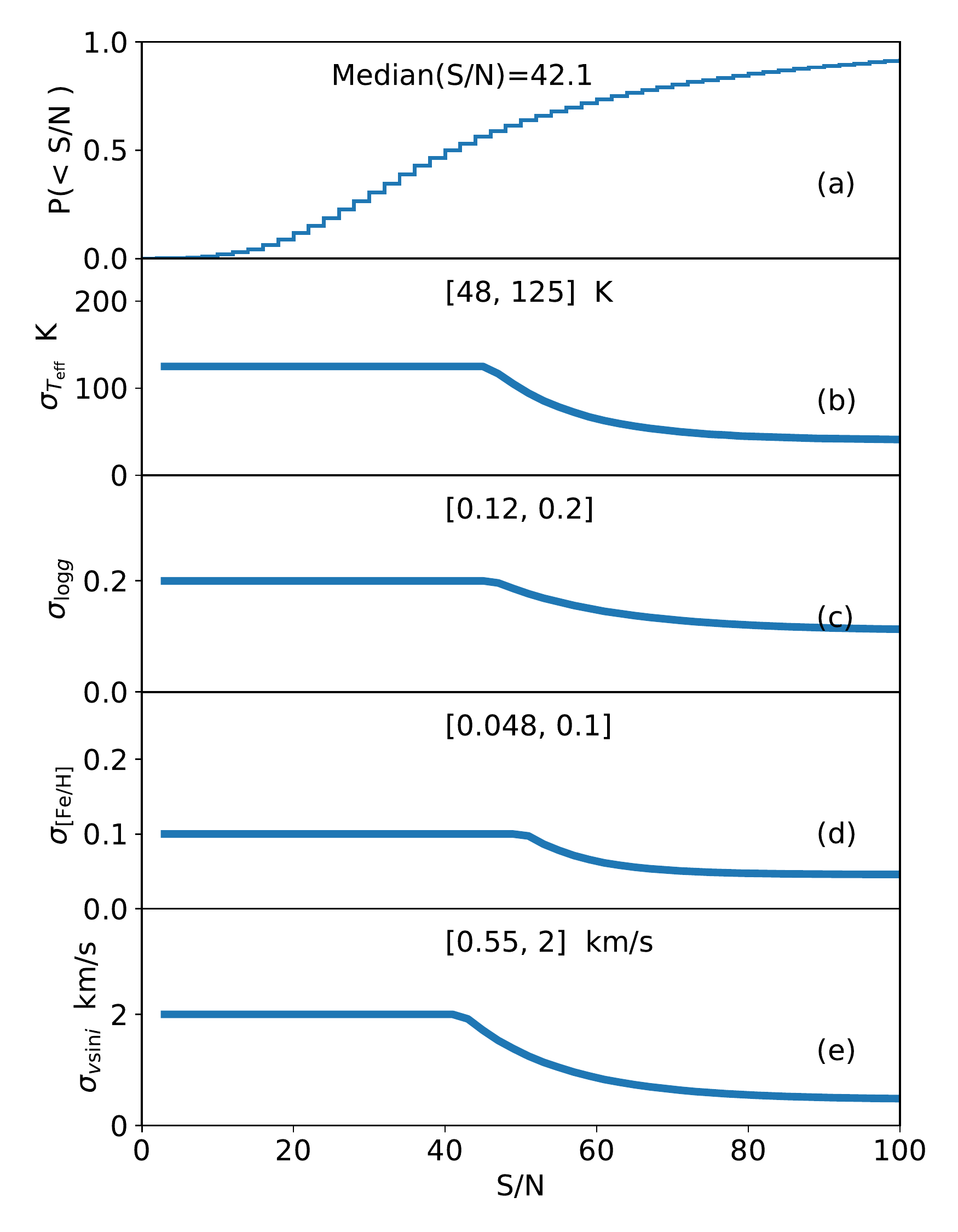}\caption{(a) Cumulative fraction of stars as function of S/N per pixel. (b)-(e) Uncertainty in spectroscopic stellar parameters as a function of S/N per pixel. The horizontal parts represent our imposed upper bounds (see text). The numbers in square bracket are the lower and upper bounds on uncertainties.
\label{fig:sigma_teff}}
\end{figure}

We use two different methods to estimate uncertainties and combine them quadratically to get the final estimate.  The resulting uncertainties are shown in \autoref{fig:sigma_teff}.
The first method is based on the covariance matrices.
The second method estimates uncertainty by fitting an exponential relation to the $rms$ of leave-20\%-out tests applied on the training data for different S/N-bins. The first method was typically found to underestimate the uncertainties but we still include it to provide a conservative lower bound on uncertainties for cases where the second method gives uncertainties that are lower than the first method.
It has to be noted that the training set only contained stars with S/N above 50. Hence, for stars with S/N below 50 we have to rely on extrapolation, which progressively becomes unreliable at lower values of S/N.  So, we placed upper bounds on the uncertainties as shown in \autoref{fig:sigma_teff}. The upper bounds were estimated as follows.  We compared the normalized distributions of each stellar parameter for samples with high S/N ($>70$) and low S/N ($<20$), respectively.  The low S/N sample was found to show flatter and wider distributions owing to the larger intrinsic uncertainty in the spectra.  We then convolved the stellar parameters in the high S/N sample with additional random uncertainty till the normalized distributions agreed with that from the low S/N sample.  These uncertainties were then adopted as the upper bounds.  We noticed that increasing these upper bounds by a factor of two, led to a noticeable mismatch in the above distribution comparisons.
For reference, we will refer to the combined data reduction and spectral analysis adopted in this work as Galah-v1.3.

\section{Derived stellar properties}\label{properties}

The problem of estimating the properties of a star given some observables can be formulated as follows. Let
\begin{eqnarray}
{\bf y}=(l',b',m_J',m_{Ks}',T_{\rm eff}',\log g',{\rm [Fe/H]}')
\end{eqnarray}
be a set of observables associated with a star and $\sigma_{\bf y}$ their uncertainties. Here, $l'$ and $b'$ are stellar Galactic position, and $m_J$ and $m_{Ks}$ are the apparent $J$ and $Ks$ band magnitudes, respectively.  Let us denote the intrinsic variables (position, metallicity, age, initial mass, distance and extinction) of a star that we are interested in by
\begin{eqnarray}
{\bf x}=(l,b,{\rm [Fe/H]},\tau,\mathcal{M},s,A_V).
\end{eqnarray}
From theoretical isochrones we can predict ${\bf y}$ given ${\bf x}$, or in other words: a function
${\bf y(x)}$ exists.  However, we are interested in the inverse problem of estimating ${\bf x}$ given ${\bf y}$.
We seek to solve this problem using a Bayesian scheme, which we will call the Bayesian Stellar Parameters estimator, or BSTEP in short.
A Bayesian introduction to solving such a problem was given by \citet{2004MNRAS.351..487P} and \citet{2005A&A...436..127J} in the context of estimating ages \citep[see also][]{2006A&A...458..609D}. The method was further improved and refined by \citet{2010MNRAS.407..339B,Binney_Burnett_2011}
and \citet{2014MNRAS.437..351B} in the context of the estimation of distances \citep[see also][]{2017ARA&A..55..213S}, with a better treatment of priors and selection
effects \citep{2012MNRAS.427.2119S,2015MNRAS.452.2960S}.  From Bayes theorem we have
\begin{eqnarray}
p({\bf x}|{\bf y},\sigma_{\bf y}) & \propto & p({\bf y}|{\bf
x},\sigma_{\bf y})p({\bf x})
\label{bayes}
\end{eqnarray}
In the following, we describe each of the terms on the right hand side of Equation \ref{bayes}, and how we specifically decided to implement each part into BSTEP for estimating the fundamental stellar parameters in the TESS-HERMES catalog.
\begin{itemize}
\item The term $p({\bf y}|{\bf x},\sigma_{\bf y})$ is the likelihood of the observed data given a set of intrinsic parameters. For Gaussian distributed uncertainties, which we assume applies here, this implies,
\be
p({\bf y}|{\bf x},\sigma_{\bf y})=\prod_i \frac{1}{\sqrt{2\pi}\sigma_{y,i}}\exp\left(\frac{-(y_i-y_i(x))^2}{2\sigma_{y,i}^2}\right),
\ee
where $i$ denotes the $i$'th observed quantity of the star.
\item The term $p({\bf x})$ specifies the prior, here coming from a Galactic population synthesis model,  which is typically of the following form:
\begin{eqnarray}
p({\bf x}) & = & p(\mathcal{M}|\tau) p(\tau)
p({\rm [Fe/H]}|\tau) p(\mu|\tau)  p(A_V|\mu),
\label{equ:dist_func}
\end{eqnarray}
where $\mathcal{M}$ is stellar mass, $\tau$ is stellar age, [Fe/H] is stellar metallicity, $\mu$ is distance modulus, and $A_V$ is the interstellar absorption.
In our approach, we use the following more simple form
\begin{eqnarray}
p({\bf x})
& = & p(\mathcal{M}) p(\tau) p({\rm [Fe/H]}) p(\mu) p(A_V),
\label{equ:dist_func2}
\end{eqnarray}
which assumes that mass, metallicity, and distance of the stars are independent of age, and that interstellar absorption is independent of distance.
For the initial mass function we adopt the \citet{2001ApJ...554.1274C} exponential form
\be
p(\mathcal{M})= 22.8978
\exp\left[-\left(\frac{716.4}{\mathcal{M}}\right)^{0.25}\right]\mathcal{M}^{-3.3}.
\ee
We assume a constant star formation rate of $p(\tau)= 1/13.18$ across all ages $0<\tau/{\rm Gyr}<13.18$, we give an equal probability of $p({\rm [Fe/H]})= 1/2.5$ to all possible metallicities in the range $0.001<Z<0.03$ ($-2.278<{\rm [Fe/H]}< 0.198$), and we assume a prior on distance from the stellar density through the following form:
\be
p(\mu) & \propto & \rho(l,b,s) s^2 \frac{d s}{d \mu}
\nonumber \\
& \propto & \rho(l,b,s) s^3.
\ee
For the stellar density, $\rho$, we assume a two-component  Galactic model consisting of an exponential stellar disc (scale  length $R_d=2.5\,$kpc, scale height $z_d=1.0$ kpc) and a  Hernquist stellar halo with scale radius $r_{\rm H}=15$ kpc  \citep{2005ApJ...635..931B,2006JPhCS..47..195N}. The fraction of  stars in the halo was assumed to be 1\%. Finally, we assume a flat prior on extinction, $p(A_V)=1/A_V^{\infty}(l,b)$.
Note, the uncertainties on spectroscopic parameters are quite small so
the exact form of adopted priors will not have much
influence on our derived results.
\end{itemize}

The position on the sky is typically known very precisely and hence their uncertainty will have negligible effect on the derived properties.  Hence, just for notational convenience we therefore drop them from here onwards (but they are included).  Isochrones provide stellar properties for a given set of intrinsic variables ${\bf x}_{\rm I}=\{{\rm[Fe/H]}, \tau, \mathcal{M}\}$, where subscript `I' denotes `Isochrone'.  So, we can write ${\bf x}=\{{\bf x}_{\rm I},\mu,A_V\}$.
In our case, ${\bf x}_{\rm I}$ is specified on a grid, which is regular in $({\rm[Fe/H]}, \tau)$, being of size $34\times177$, but irregular in $\mathcal{M}$.  Our grid of stellar models was constructed using CMD 3.0 (\href{http://stev.oapd.inaf.it/cmd}{http://stev.oapd.inaf.it/cmd}), with  PARSEC-v1.2S isochrones \citep{2012MNRAS.427..127B, 2014MNRAS.445.4287T, 2014MNRAS.444.2525C, 2015MNRAS.452.1068C}, the NBC version of bolometric corrections  \citep{2014MNRAS.444.2525C}, and assuming Reimers mass loss with efficiency $\eta=0.2$ for RGB stars.
By interpolation, we resample the $\mathcal{M}$ dimension (for each given ${\rm[Fe/H]}$ and $\tau$) in the grid, such that $\Delta \log T_{\rm eff}< 0.004$ and $\Delta \log g< 0.01$,
which is of the order of the expected uncertainties in $T_{\rm eff}$ of 50 K from spectroscopy and in surface gravity of 2.5\% from asteroseismology, assuming the temperature is known.  After this resampling the total number of grid points were $\sim 6 \times 10^6$.

Now, if we let $x_{i,j}$ denote the $j$-th parameter (such as mass, metallicity, age, distance, and extinction) of the $i$-th grid point, the multi-dimensional volume represented by the grid point is given by
\be
V_i = \int {\rm d}^5{\bf x} = \prod_{j=1}^{5} \frac{(x_{i,j+1}-x_{i,j-1})}{2}.
\ee
The five dimensions that span the volume are [Fe/H], $\tau$, $\mathcal{M}$, $\mu$, and $A_V$. Finally, we can derive the probability of the fundamental stellar properties given the observations, as
\be
p({\bf x}|{\bf y},\sigma_{\bf y}) =
\int p({\bf
y}|{\bf x},\sigma_{\bf y}) p({\bf x}) {\rm d}^5{\bf x}
\nonumber \\
=  p({\bf y}|{\bf x}_i,\sigma_{\bf y}) p({\bf  x}_i) V_i
\ee

Sampling the full five dimensional space poses computational problems. So, for each star we first identify the set of grid points that are within 5$\sigma$ of the following observables: [Fe/H], \teff, and $\log g$ (\numax, and \dnu\ could also be used if available, which they are not for this paper).

In principle, BSTEP can use apparent magnitudes to further help constrain the search volume of the grid.
However, in addition to photometric apparent magnitudes, $m_{\lambda}$, ($\lambda$ being the central wavelength of the photometric band) we would also require distance, $s$, and extinction, $A_V$, following
\be
m_{\lambda} & = & M_{\lambda}+\mu+A_V f_{\lambda},
\ee
in order to relate $m_{\lambda}$ to the associated absolute magnitude, $M_{\lambda}$. Here, $\mu=5 \log (100 s/{\rm kpc})$ is the distance modulus and $f_{\lambda}=A_{\lambda}/A_V$ is the coefficient of extinction for the photometric band centered on wavelength $\lambda$.
For $f_{\lambda}$, we use Table 6 of \citet{1998ApJ...500..525S}.
The required additional input, $\mu$ and $A_V$, are not available for most of our stars.  Hence, for this paper we consider all possible values,  $A_V = \{0,...,A_V^{\infty}\}$ and
$\mu = \{0,...,m_{\lambda}+10\}$ (assuming the brightest star to have an absolute magnitude of $M_{\lambda}=-10$).
However, given a color and an apparent magnitude,
the full $(A_V,\mu)$ space does not need to be explored.  Hence, we design a scheme based on so-called importance sampling to explore the $(A_V,\mu)$ space efficiently.
For a given $\{{\rm[Fe/H]}, \tau, \mathcal{M}\}$, we sample $A_V$ in accordance with the probability distribution of the observed stellar color. Similarly, given $\{{\rm[Fe/H]}, \tau, \mathcal{M}, A_V\}$, we sample $\mu$ in accordance with the probability distribution of the observed apparent magnitude.

Out of all stars in our sample with spectroscopic parameters, about 200 did not return results from BSTEP because the input parameters fell outside the stellar model grid of BSTEP.

\section{The Catalog}\label{catalog}

In this section we describe the stars and their parameters that constitute the current data release.  This release is mainly focused on dwarfs/subgiants hotter than $4800\,$K, for which the published catalog contains an extensive list of spectroscopic results (Section \ref{spec_anal}) as well as inferred stellar properties from our BSTEP isochrone grid modeling (Section \ref{properties}).  In the following, we will refer to these stars as the hot dwarfs.  For cooler dwarfs we list mass and radius, and for giants we list $\log g$ and radius.  The separation between these three groups is defined in the spectroscopic Hertzsprung-Russell (HR) diagram as described below.

\begin{figure}
\centering \includegraphics[width=0.49\textwidth]{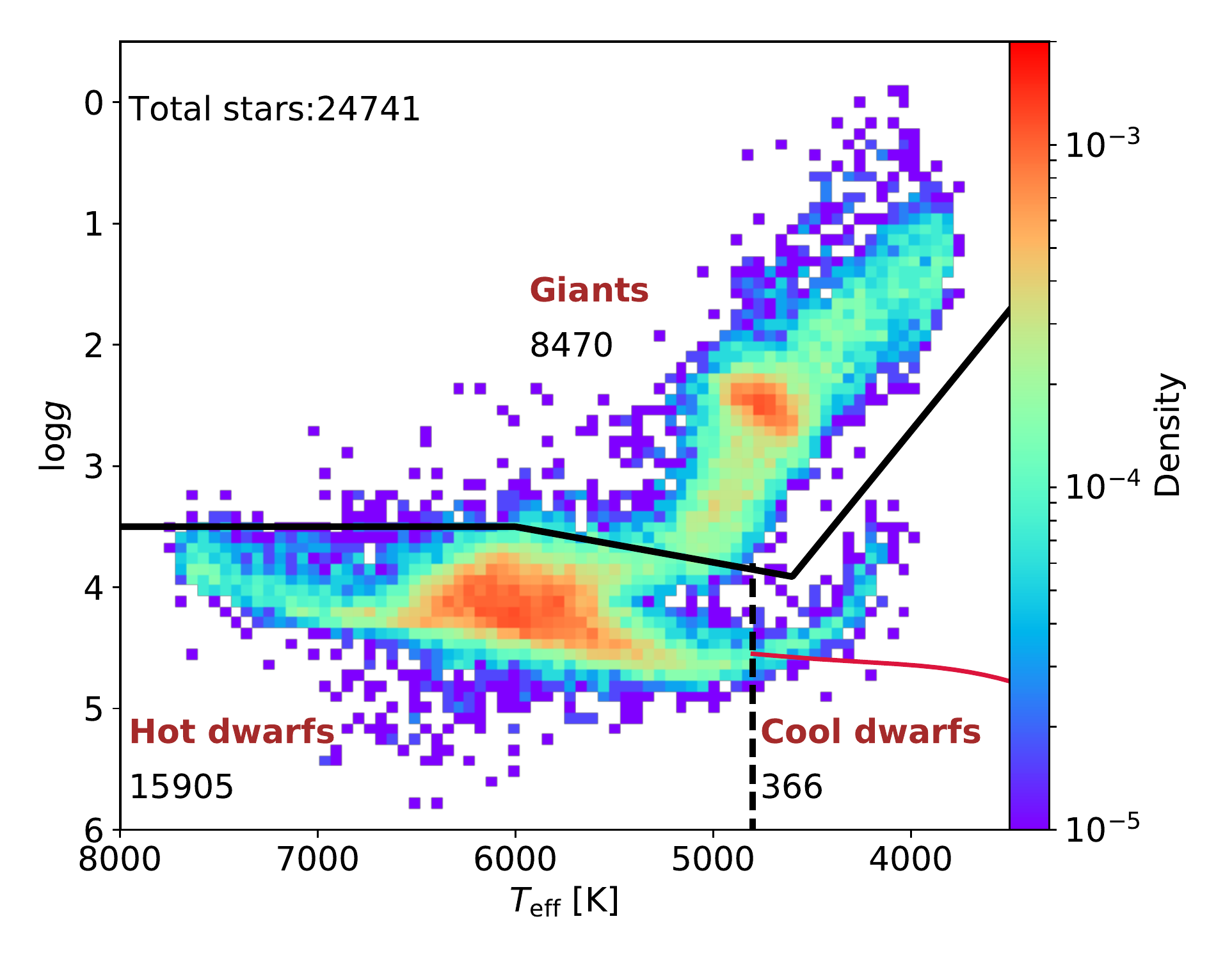}\caption{Distribution of TESS-HERMES stars in the $(T_{\rm eff},\log g)$ plane.  The solid line shows the relation used to separate dwarfs from giants. The vertical dashed line is used to separate hot and cool dwarfs ($T_{\rm eff}=4800\,$K). The red curve shows the mass-radius-\teff\ relation for the cool dwarfs adopted by \citet[][Table 3 and 4 therein]{2017arXiv170600495S}.
In this plot we excluded stars, which we identify as belonging to the LMC (see \autoref{lmc}).
\label{fig:teff_logg2}}
\end{figure}

In \autoref{fig:teff_logg2} we show an HR diagram (\teff-$\log g$ plane) for all observed stars, except the Large Magellanic Cloud (LMC) sample (see Section  \ref{lmc}).
The main stellar populations, the main sequence, the subgiants, the red giant branch, and the red clump can clearly be seen and follow the expected trends.
However, the cool dwarfs (\teff\ $< 4800\,$K) show an unphysical upturn towards lower values of $\log g$ and \teff, which we attribute to limitations in the 1D modeling of the stellar atmospheres adopted in our spectroscopic analysis.  We deem the spectroscopic stellar parameters unreliable for these stars, and for the current data release we do therefore not provide their spectroscopic parameters or the associated stellar properties from BSTEP.
Despite this `artificial' upturn, the spectroscopy still provides an unambiguous classification into either dwarfs or giants.  For the affected dwarfs (defined by $T_{\rm eff}< 4800\,$K; \autoref{fig:teff_logg2} dashed line), we couple their spectroscopic classification with the suggested mass- and radius-$T_{\rm eff}$ relations for main sequence stars from the TIC to obtain the stellar mass and radius \citep[][Table 3 and 4 therein]{2017arXiv170600495S}.  These masses and radii should be the same as in the TIC for stars that are truly dwarfs, but different for stars that are incorrectly tagged as giants in the TIC.

For the purpose of determining which stars are dwarfs or giants in the catalog we use the following $\log g$ relation (\autoref{fig:teff_logg2}, solid line),
\begin{equation}
\log g= \begin{cases}
3.5, &\text{for\ } T_{\rm eff}/{\rm K} > 6000 \\
3.912-(T_{\rm eff}/{\rm K}-4600)\frac{0.412}{1400}, &\text{for\ } 4600<T_{\rm eff}/{\rm K} < 6000 \\
3.912+(T_{\rm eff}/{\rm K}-4600)\frac{5.2}{2600}, &\text{for\ } T_{\rm eff}/{\rm K} < 4600
\end{cases}
\end{equation}
This relation is similar to \citet{2011AJ....141..108C} but differs slightly for $T_{\rm eff}<6000\,$K.  The low temperature ($T_{\rm eff}<4600\,$K) regime of the relation is specifically designed to separate the cool dwarfs from giants in our spectroscopic sample.
For 43 cool dwarfs we have parallax information available from Gaia DR1 and we find that all of these stars have absolute magnitude
$M_J>4.2$, confirming that they are indeed dwarfs.

\subsection{The Large Magellanic Cloud}\label{lmc}

\begin{figure}
\centering \includegraphics[width=0.48\textwidth]{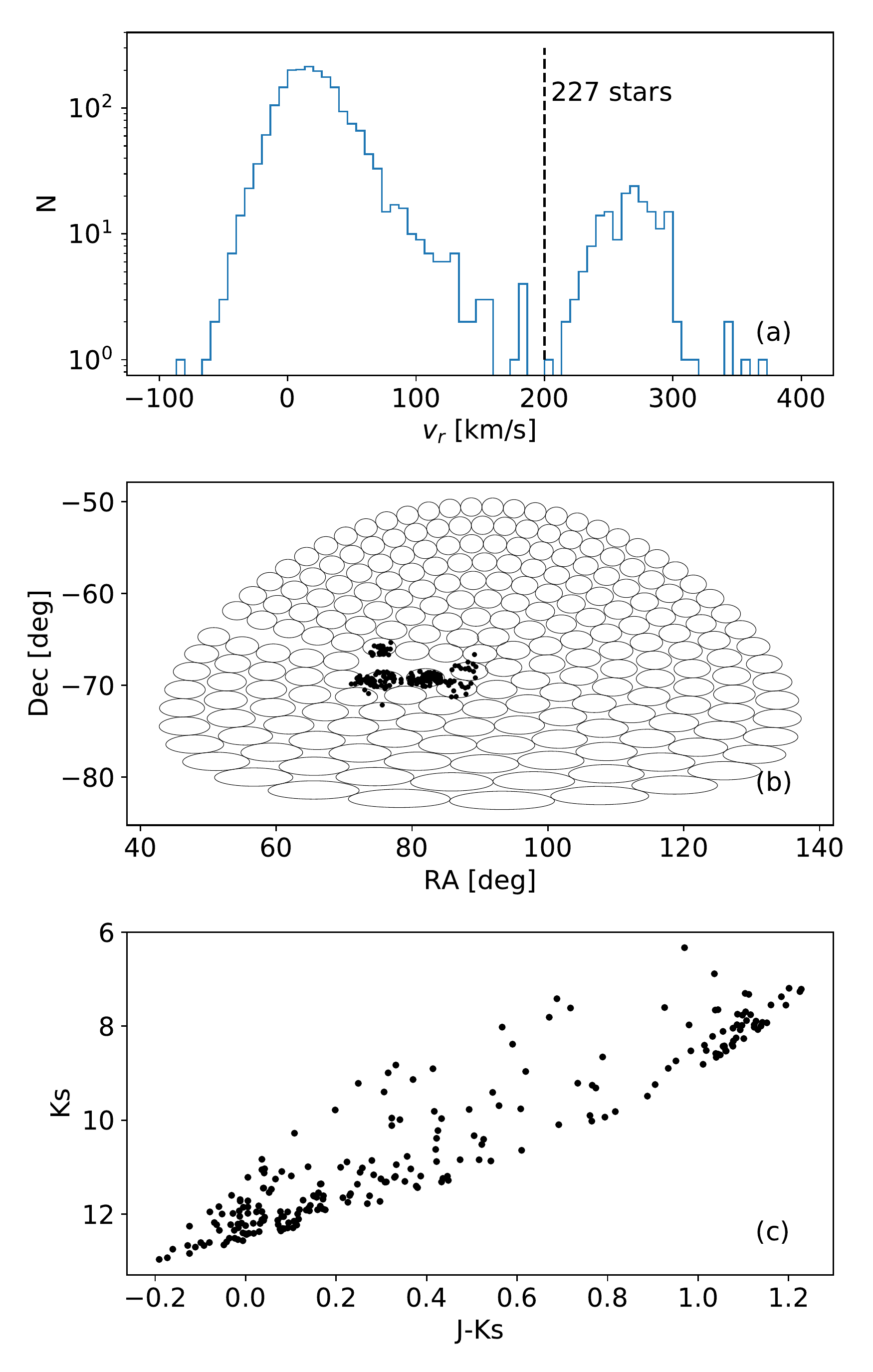}\caption{(a) The distribution of heliocentric velocity of all observed stars in our sample. The dashed line marks the threshold used to identify stars that belong to the LMC beyond 200 km/s. We also have LMC stars with $v_{\rm helio}>400\, {\rm km/s}$, whose velocities are spuriously elevated due to strong emission features
in the spectra.
(b) Location of identified LMC stars (bold points) relative to all fields of the TESS-HERMES survey. (c) Color magnitude diagram of identified LMC stars.
\label{fig:tess_lmc1}}
\end{figure}

The LMC lies in the southern TESS CVZ.  Although it is 49.5 kpc away, its brightest stars (e.g. Wolf Rayet stars, hot O and B type stars, and cool supergiants) have apparent magnitudes that fall within our observable range.  The spectrum of some of these stars show strong emission lines in H$\alpha$ and He.  We also see stars with P Cyg profiles.  Our current spectroscopic pipeline for estimating stellar parameter is not tuned to deliver reliable stellar parameters for stars with such peculiar spectral features. However, for most of them the radial velocity is  reliably determined by GUESS. The distribution of radial velocity of our stars clearly shows a bimodal distribution (\autoref{fig:tess_lmc1}a).
The first mode is due to stars in the Milky Way. The second mode, centered around 270 km/s, coincides with the velocity of the LMC (262 km/s).  The line with $v_{\rm helio}=200$ km/s cleanly separates the two distributions.  Hence, to identify the LMC stars we devise the following criteria, $v_{\rm helio} > 200 {\rm\, km/s}$ and $70^{\degree}<{\rm RA}<90^{\degree}$ and $-72.4^{\circ}<{\rm Dec}<-65^{\circ}$.  This yielded 227 LMC stars, which form a distinct group on sky (\autoref{fig:tess_lmc1}b).  The color magnitude diagram (\autoref{fig:tess_lmc1}c) shows that the stars are either hot stars or cold stars, with very few in between.  The correlation of color with magnitude is because of our $V_{JK}<13.1$ selection function, which leads to a color dependent upper limit on $Ks$. In the catalog,  we mark these stars as being part of the LMC, but leave their stellar parameters as undefined, and do not include them in the following analysis of this paper.

\subsection{Classification of stars into dwarfs, subgiants and giants}\label{separation}

\begin{figure}
\centering \includegraphics[width=0.48\textwidth]{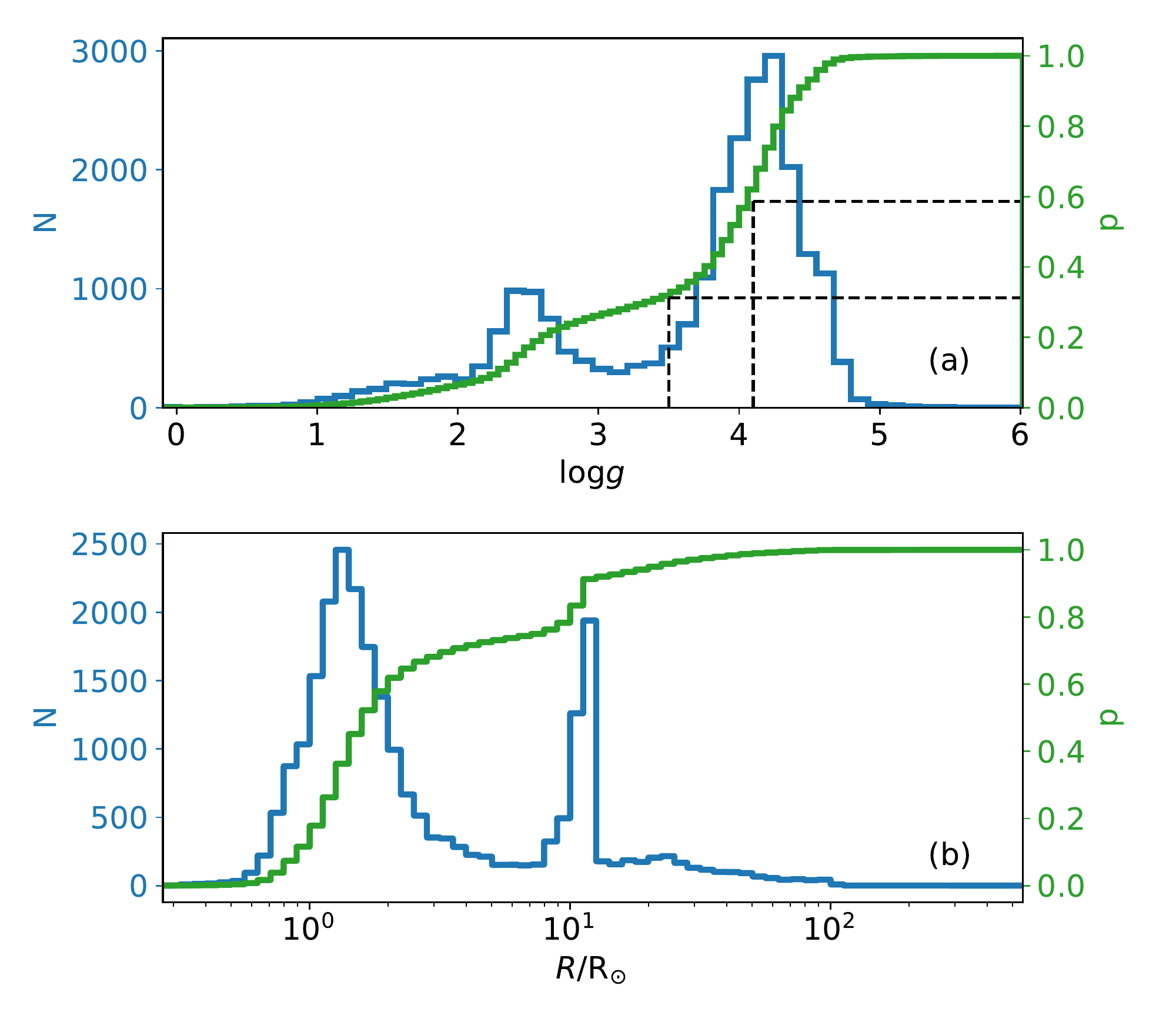}\caption{Differential and cumulative distributions of spectroscopic-based (a) surface gravity and (b) radius of all observed TESS-HERMES stars.
\label{fig:tess_logg_r}}
\end{figure}

The primary purpose of the TIC is to support the selection of roughly 400,000 top priority stars for the TESS 2-minute cadence observations, which will form the mission's primary data for the search of small ($<2.5 R_{\oplus}$) transiting planets.  For this purpose, the TIC defines dwarfs as stars with $\log g>4.1$, subgiants as $3.5<\log g<4.1$ and giants as $\log g <3.5$.   Adopting these definitions, we find, from the $\log g$ distribution of all the TESS-HERMES stars observed so far, that about $40\%$ are dwarfs, $30\%$ are subgiants, and $30\%$ are giants (\autoref{fig:tess_logg_r}a, see dashed lines).  For completeness we also show the sample's radius distribution (Figure \autoref{fig:tess_logg_r}b).

\begin{figure}
\centering \includegraphics[width=0.48\textwidth]{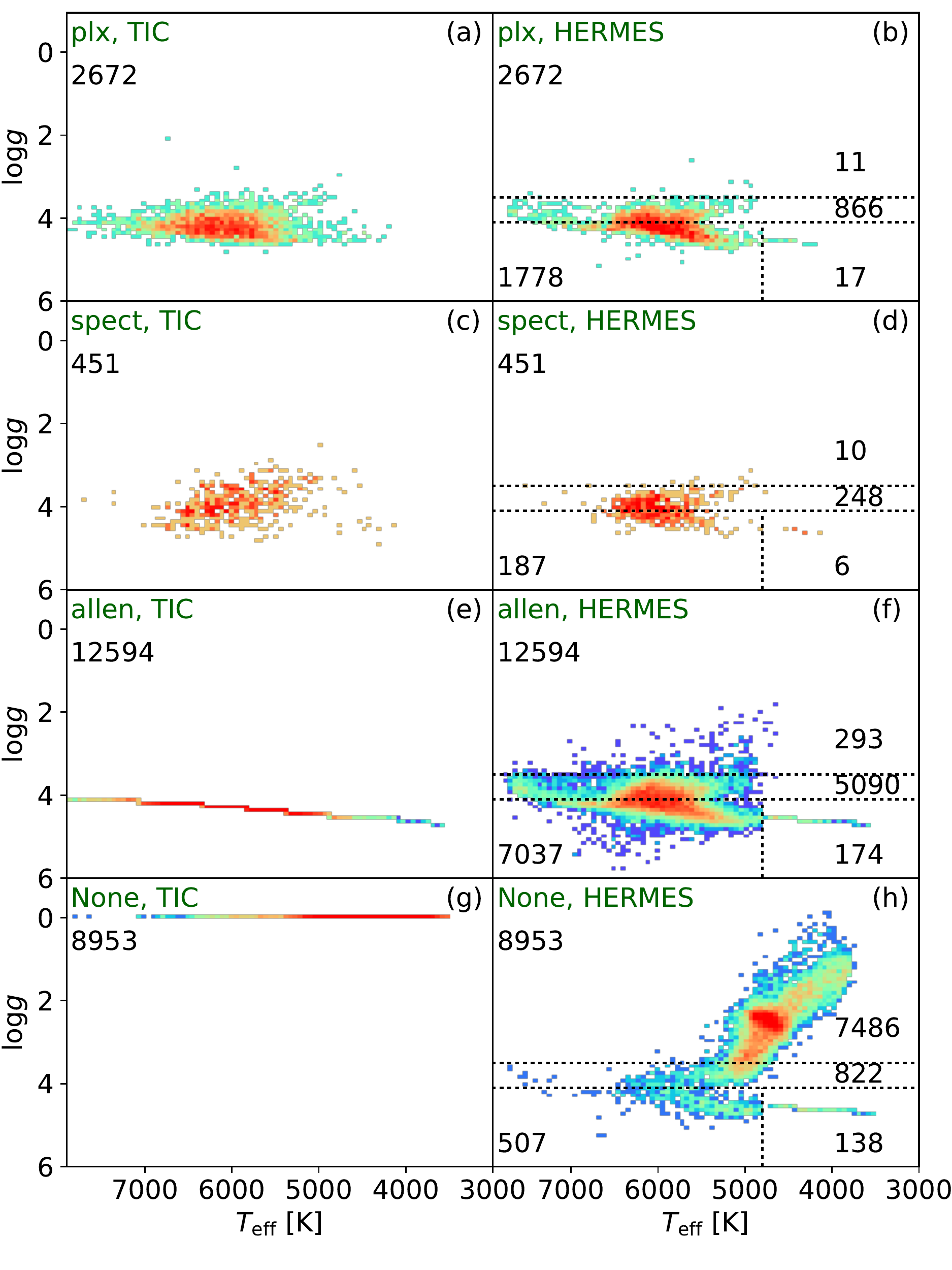}\caption{The HR diagram for the observed TESS-HERMES sample using stellar parameters from the TIC (left panels) and parameters derived using the HERMES spectra (right panels).  Each row of panels represent one approach used in the TIC to determine the stellar parameters.  The source of the stellar parameter determination in the TIC (given by the TIC spflag: plx, spect, allen, or None) for dwarfs and giants (meaning stars with TIC lumclass not equal to None) is listed in the top left in each panel.  The number in the top left shows the number of stars in the panel.  The numbers in the lower half of the right panels are star counts in each part of the panels separated by the dashed lines, which represents (clock-wise), giants, subgiants, cool dwarfs, and hot dwarfs, following the TIC $\log g$ separations of giants, subgiants, and dwarfs. The color coding is indicative of the density of data points.
\label{fig:tic_comp_cases}}
\end{figure}

In the following we want to compare our spectroscopic results with what is listed in the TIC for the TESS-HERMES sample. We match stars using the 2MASS identifier. In the TIC, the stellar parameters are determined using different approaches depending on what information is available for a given star.
There are four main approaches, denoted by the `spflag' in the TIC as either: plx, spect, allen, or None.  We show the HR diagram for the stars associated with each approach in \autoref{fig:tic_comp_cases} (left panels).
We now discuss each of them in detail.
\begin{itemize}
\item plx: This is the approach used when the parallax information is available from Hipparcos \citep{2007A&A...474..653V} or Gaia DR1 \citep{2016A&A...595A...1G, 2016A&A...595A...2G, 2016A&A...595A...4L}. Parallax and temperature is used to estimate luminosity and radius.  The mass is also estimated from the temperature.  The HR diagram of the TIC parameters (\autoref{fig:tic_comp_cases}a) matches nicely those based on the TESS-HERMES spectroscopy (\autoref{fig:tic_comp_cases}b)\footnote{We note that the $\log g$ values in the TIC (column 67) showed a much larger spread than in \autoref{fig:tic_comp_cases}a. We therefore derived $\log g$ from mass (column 71) and radius (column 73).}.
\begin{figure}
\centering \includegraphics[width=0.48\textwidth]{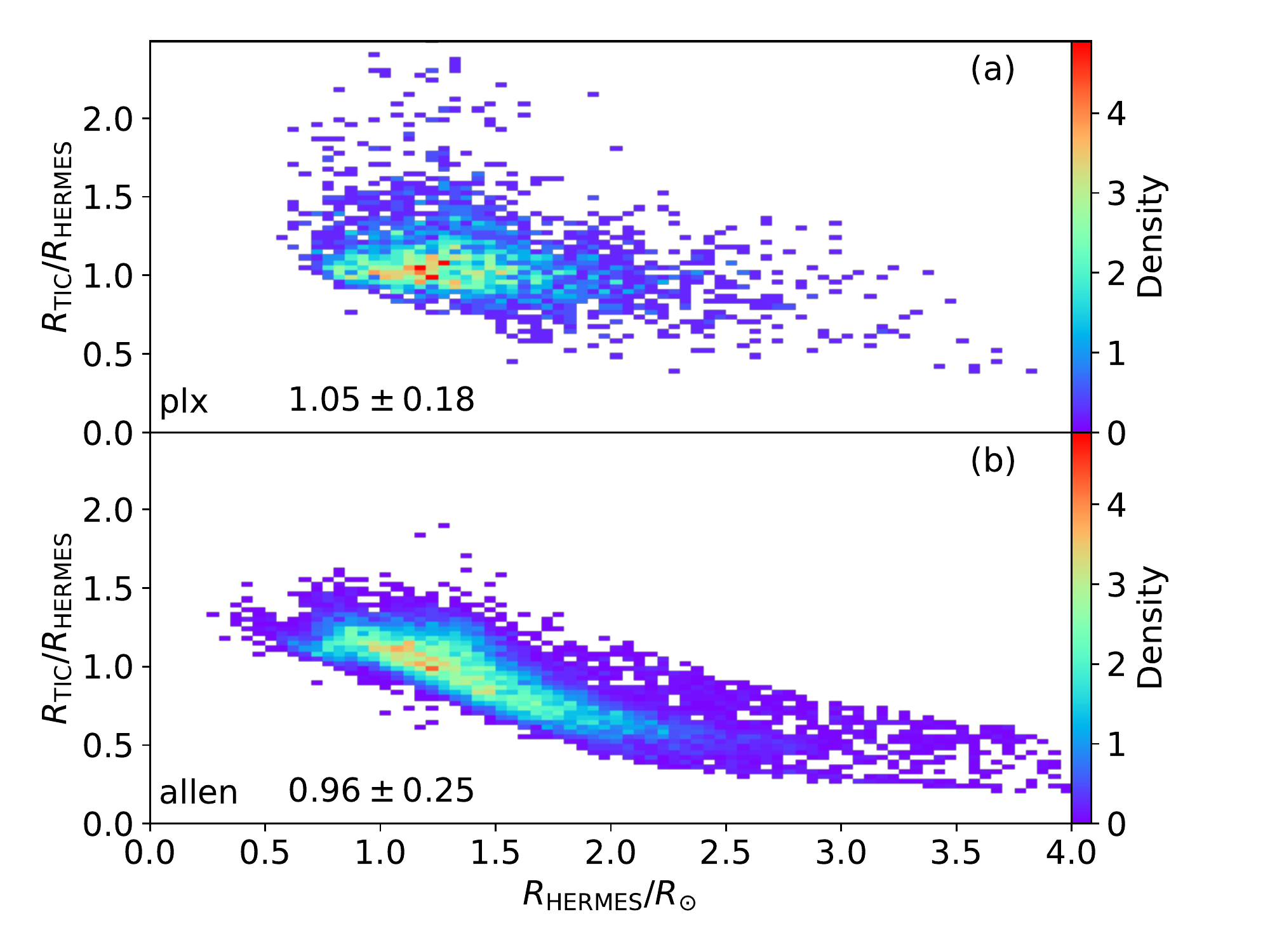}\caption{Comparison of TIC radius estimates with those determined from the TESS-HERMES spectroscopy. (a) For the case where parallax information is available (TIC spflag = plx). (b) For the case where only photometric information is available (spflag = allen). The median TIC/HERMES radius ratios are printed in the lower left. \label{fig:tic_comp_rad}}
\end{figure}
In \autoref{fig:tic_comp_rad}a, we compare the radius estimates for the `plx' sample, and find that the TIC estimates agree well with our spectroscopically determined results.  However, this sample only probes stars of similar size to the Sun, and we do not know if the TIC radius estimates will behave similarly for more evolved stars.
\item spect: This is the approach used when spectroscopic information is available in the TIC. The spectroscopic information in TIC is from SPOCS \citep{2016ApJS..225...32B}, PASTEL \citep{2016A&A...591A.118S},  Gaia-ESO \citep{2012Msngr.147...25G}, GALAH \citep{2017MNRAS.465.3203M}, APOGEE \citep{2015AJ....150..148H}, LAMOST \citep{2015RAA....15.1095L} RAVE \citep{2013AJ....146..134K} and GCS \citep{2009A&A...501..941H} surveys.
This constitutes a very small fraction of stars in our sample.  Our TESS-HERMES results (\autoref{fig:tic_comp_cases}d) scatter much less than the spectroscopy results in the TIC (\autoref{fig:tic_comp_cases}c).
\item allen: This is the adopted approach when only photometric and proper motion information is available, and the stars are classified as dwarfs using reduced proper motion.  This category constitutes the majority of the dwarf sample in the TIC.  The mass, radius, and gravity are assumed to be simple analytical functions of temperature.  This manifests itself as a simple curve in \autoref{fig:tic_comp_cases}e.  With the TESS-HERMES spectra, we can now separate dwarfs from subgiants and giants (\autoref{fig:tic_comp_cases}f).  We find that 43\% of this subsample are subgiants and giants [(293+5090)/12594].  In \autoref{fig:tic_comp_rad}b, we compare our TESS-HERMES radius estimates with those from the TIC, and find that the TIC estimates agree well for stars of similar radius to the Sun, but underestimates radius for more evolved (larger) stars.  This is a result of forcing the subgiants onto the main sequence relations in the TIC.
\item None: This is the approach used when only photometry and proper motion is available, but no estimate of radius, mass, and $\log g$ is listed in the TIC (values are NaN in the TIC).  For illustration in \autoref{fig:tic_comp_cases}g, we have set their $\log g$ to zero.  Almost all of these stars are expected to be giants (TIC lumclass = GIANT)\footnote{There are 54 stars in our sample that are unclassified in the TIC (lumclass=None), most likely due to unreliable proper motions, and we excluded them from our analysis.}.  Our results show that most of the stars are indeed giants (\autoref{fig:tic_comp_cases}h).  However, about 7\% [(507+138)/8953] of the stars are in fact dwarfs.  More importantly, the sample contains 138 cool dwarfs, which is about 40\% of the total number of 366 cool dwarfs in our current TESS-HERMES sample.  These should be high priority targets for exoplanet searches.
\end{itemize}

\begin{figure}
\centering \includegraphics[width=0.48\textwidth]{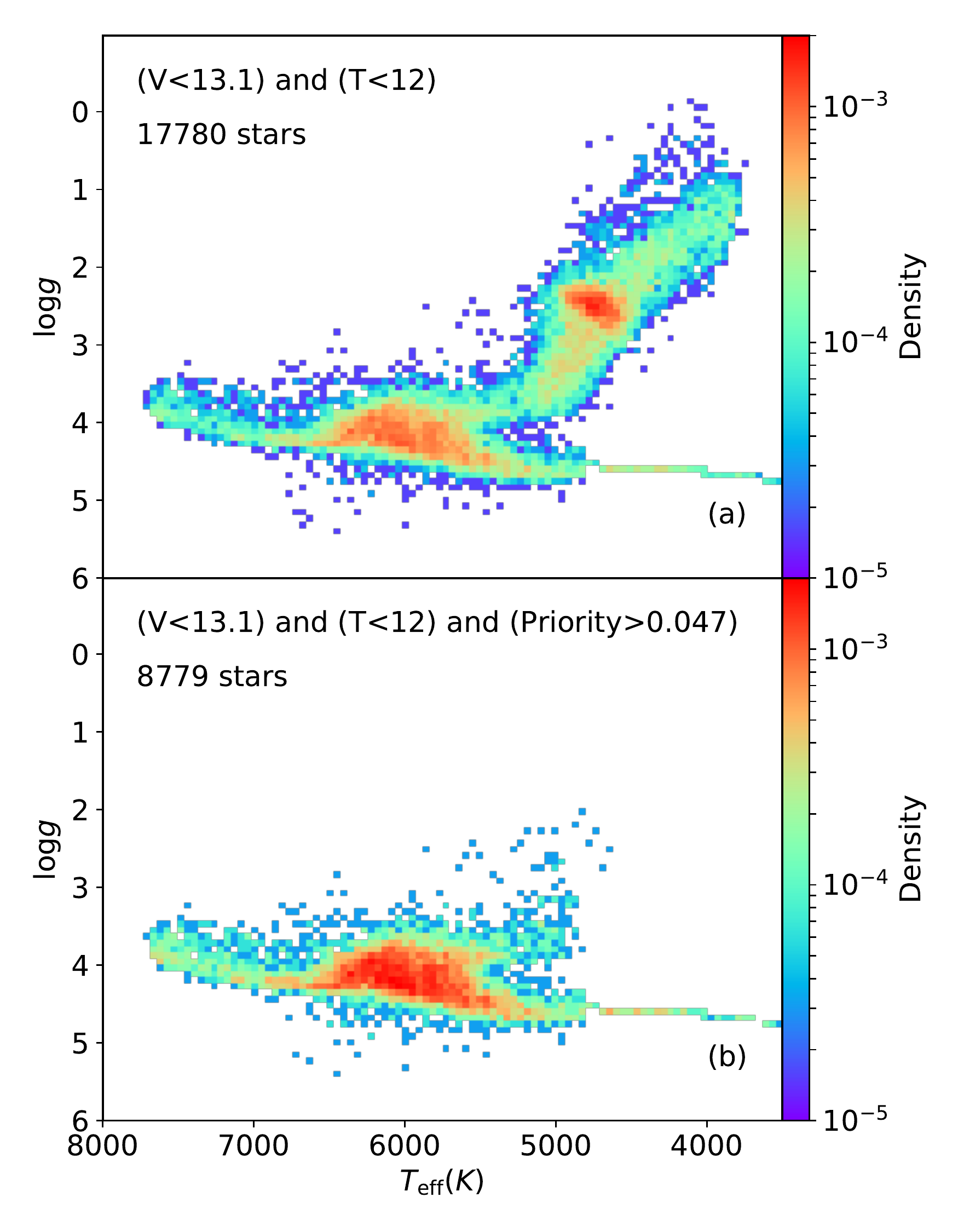}\caption{(a) HR diagram from spectroscopy of the observed TESS-HERMES sample restricted to the brightest stars ($T<12$).  (b) As panel (a), but restricting the sample further to those with a priority in the CTL of $>0.047$.
\label{fig:tic_comp4}}
\end{figure}

For the purpose of finding small transiting planets, the TIC assigns a priority to a star that is inversely proportional to its radius.  Based on this, a candidate target list (CTL) of stars sorted according to priority is constructed.  The top 400,000 stars in this list have priority $>0.047$.  Of these, 90733 lie within 13 degrees of the southern ecliptic pole, of which 21,848 are brighter than 12 mag in $T$.  As discussed in \autoref{sec:observations}, our TESS-HERMES sample is unbiased for $T<12$.  The HR diagram of our observed stars with $T<12$ is shown in \autoref{fig:tic_comp4}a.  We have stellar parameters for 8779 of the 21,848 bright high priority targets and the HR diagram of these is shown in \autoref{fig:tic_comp4}b.
It can be seen that the approaches used in the TIC for stellar parameter determination have successfully flagged almost all giants with $\log g< 3.5$ (those not suitable for detecting transits of small planets).  However as discussed above, the TIC, and hence the CTL, misses out on a significant fraction of bright cool dwarfs, which we are now able to highlight with our survey.  In fact, a total of 149 cool dwarfs with $T_{\rm eff}<4800\,$K and $10<V_{JK}<13.1$ lie within 13 degree of the ecliptic pole and have priority less than 0.047. Three of them are classified as cool dwarf in the TIC but have a priority of zero, while 8 of them are only just outside the 12 degree circle around the ecliptic pole.

In \autoref{tab:classerror} we list the expected misidentification and contamination percentages in the TIC.  The type of error in classifying the stars is taken from \citet{2017arXiv170600495S} (their Fig. 5).  We compare these expected errors with the `true' errors based on our TESS-HERMES spectroscopic results. The TIC classifications are based on the TIC $\log g$ values.  Dwarfs are stars with $\log g>4.1$ and TIC lumclass='DWARF', subgiants have $3.5<\log g<4.1$ and lumclass='DWARF', and giants have $\log g <3.5$ and lumclass='GIANT'.  There is generally very good agreement between the expected and the spectroscopic-based results.  However, naively our estimate of 4.5\% for "dwarf contaminants among giants in TIC" (\autoref{tab:classerror}), is higher than that reported by \citet{2017arXiv170600495S} for the TIC (their Figure 5).
This is due to a number of factors.
First, the error rates will depend upon color magnitude limits of the spectroscopic sample, and our sample is very different from the one used in the TIC. Second we have excluded stars with parallax information, which will bias
our results. Third, the quoted errors in the TIC are for stars with proper motions from Gaia DR1, while the majority of our stars have proper motions from UCAC4.  In general, the rate depends upon the source of proper motion used in the TIC, and we confirmed that UCAC4 gives a higher error rate than Gaia DR1.  Moreover, the proper motion uncertainty in UCAC4 is higher in southern declinations, which is where our survey is. For similar reasons, our rate for "dwarfs misidentified as giants in TIC" is also slightly higher than in the TIC. After the release of Gaia DR2, it is expected that the updated TIC will have contamination rates as quoted in TIC. However, we find that 6.4\% of dwarfs identified by us have undefined radius in the TIC (and hence a priority of zero) (\autoref{tab:classerror}). Most of these do have a dwarf and giant classification in the TIC (lumclass) based on reduced proper motion,  with 56\% of having TIC lumclass='GIANT' and 42\% having lumclass='DWARF'.

\begin{table}
\caption{Classification errors in the TIC}
\begin{tabular}{@{}|l|l|l|}
\hline
Error type & Expected & True  \\
\hline
Dwarfs misidentified as giants in TIC & 2.0\% & 4.8  \% \\
Giants misidentified as dwarfs in TIC & 4.0\% & 3.7  \% \\
Subgiants misidentified as dwarfs in TIC    & 85\%  & 84.  \% \\
Dwarf contaminants among giants in TIC      & 1.0\% & 4.5  \% \\
Giant contaminants among dwarfs in TIC      & 2.5\% & 2.3  \% \\
Subgiant contaminants among dwarfs in TIC   & 50\%  & 40.  \%  \\
\hline
Dwarfs with undefined radius in TIC &  &  6.6\% \\
\hline
\end{tabular}
\label{tab:classerror}
\tablecomments{The `Expected' errors are taken from \citet{2017arXiv170600495S} (their Fig. 5).  The `True' errors are based on the TESS-HERMES spectroscopy, and are computed for samples with the following restrictive flags in the TIC: (spflag $==$'allen') OR (spflag $==$'None').}
\end{table}

Currently there are about 90,000 high priority targets in the CVZ region, while the expected number of available 2-minute cadence slots is only about 10,000.  Hence, it will be important to de-prioritize stars with larger radii based on accurate radius estimates in order to optimize the exoplanet target list.  With our current observations, we have already identified 5573 stars that lie within 13 degrees of the ecliptic pole, which have priority$>0.047$, and $\log g > 4.1$.

\subsection{Comparison between TIC and TESS-HERMES effective temperatures}

\begin{figure}
\centering \includegraphics[width=0.48\textwidth]{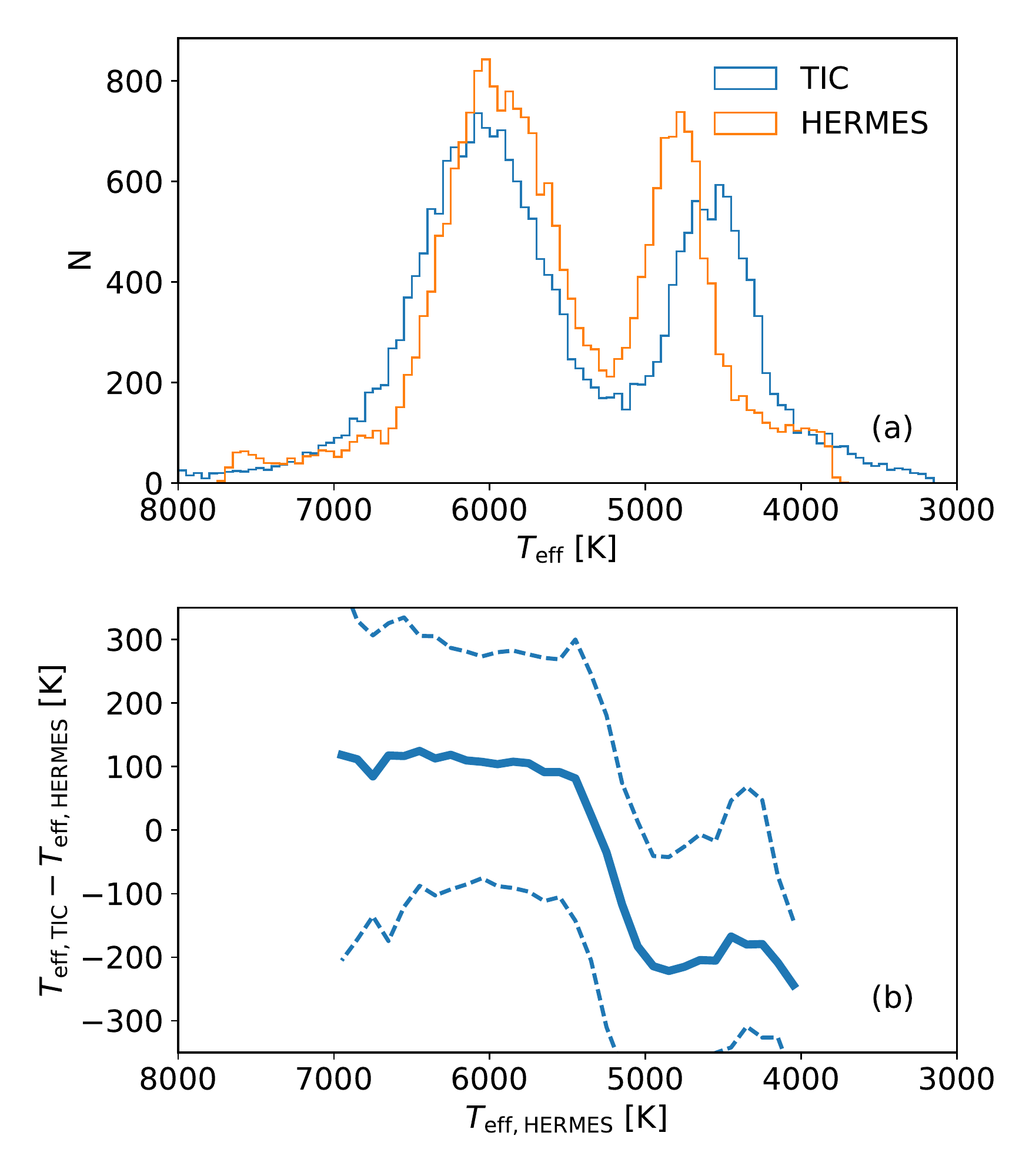}\caption{(a) Distributions of \teff\ from the TIC and the HERMES spectra (hot dwarfs and giant sample). (b) Like (a), but shown as the difference in \teff\ between TIC and HERMES as a function of \teff.  The solid line shows the 50th percentile, while the dotted lines show the 16th and 84th percentiles.
\label{fig:tic_teff_comp}}
\end{figure}
In \autoref{fig:tic_teff_comp}, we compare the TIC effective temperatures with those derived from the HERMES spectra in this study (for hot dwarfs and giants).  The distribution shows two peaks, the hotter corresponding to main sequence stars and the cooler corresponding to giants.  For main sequence stars, the TIC \teff\ is about 100\,K hotter compared to the spectroscopic estimates.  We found a roughly $200\,$K difference when comparing the distribution of spectroscopic \teff\ to that of the Galactic model, {\sl Galaxia}, which, was hotter and like the TIC, bases its \teff\ on photometry.  This discrepancy might be related to differences between the spectroscopic and photometric temperature scales \citep{2012ApJS..199...30P}.
For giants, the TIC $T_{\rm eff}$ is cooler by about $200\,$K.  The TIC \teff\ scale for stars with no spectroscopy is based on the calibrations of \citet{2015MNRAS.454.2863H}, who reports their scale is about $100\,$K cooler than that based on MARCS model atmospheres \citep{2008A&A...486..951G}.  Hence, this may account for half of the difference seen here.  The other half could be due to underestimation of extinction for giants in the TIC.

\subsection{Distribution of stellar properties in the Catalog}

\begin{figure}
\centering \includegraphics[width=0.48\textwidth] {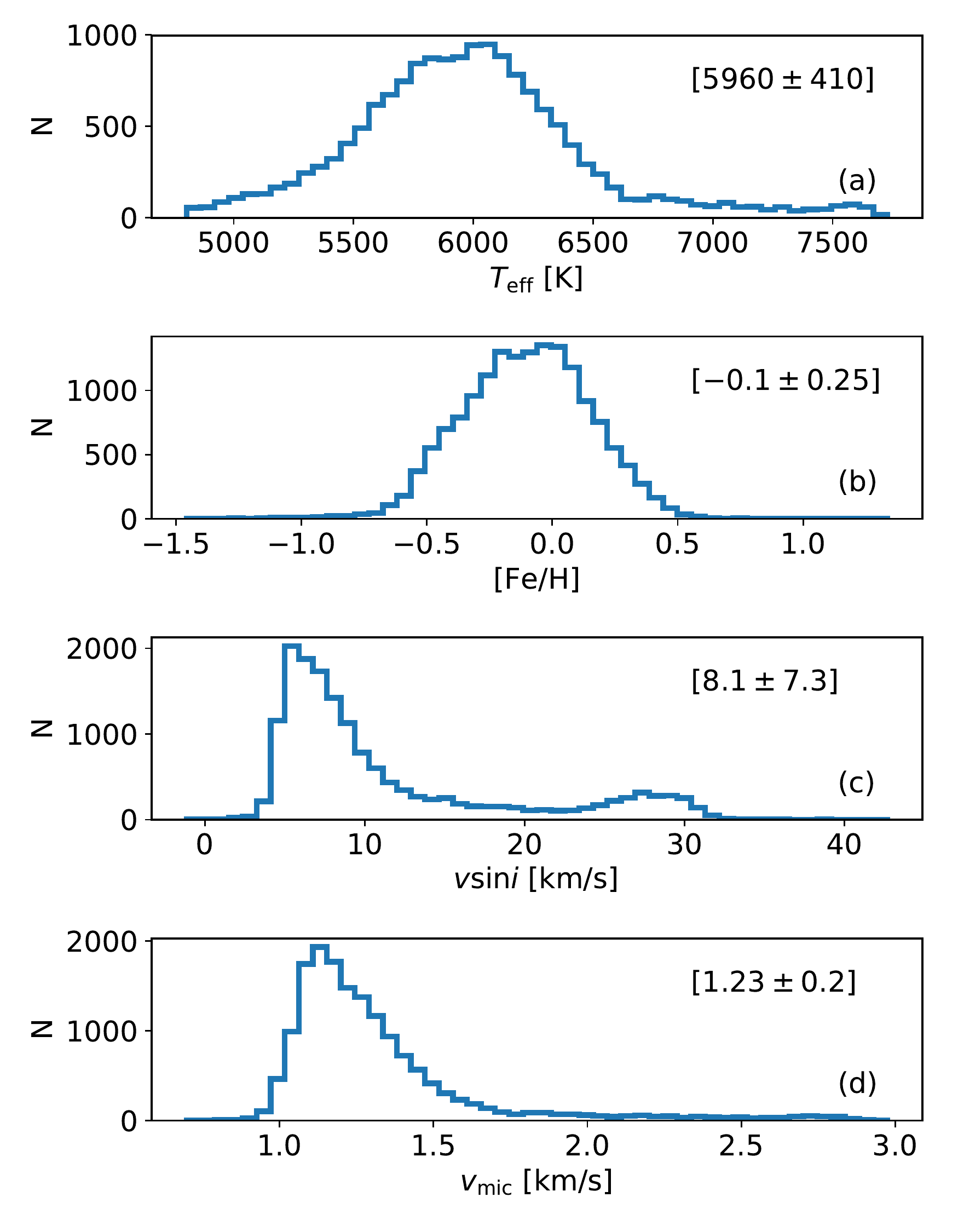}\caption{Distribution of spectroscopic parameters, temperature, metallicity, rotational, and micro-turbulence velocity for the observed `hot dwarf sample'. Median and standard deviation values of each parameter are shown in the top right of each panel.
\label{fig:tess_teff_feh_v}}
\end{figure}

We now turn our focus entirely to the hot dwarfs, which are the main focus of the current data release ($T_{\rm eff}> 4800\,$K, lower left group in \autoref{fig:teff_logg2}). The distribution of their spectroscopic parameters ($T_{\rm eff}, {\rm [Fe/H]}, v \sin i, v_{\rm mic}$) is shown in \autoref{fig:tess_teff_feh_v}.
With \teff\ peaking around $6000\,$K and [Fe/H] around -0.1 dex, it is evident that the TESS-HERMES sample is well suited for selecting Sun like stars.  For the majority of stars the rotational velocity ($v \sin i$) is less than 10 km/s, which suggests that the stars are generally expected to be good candidates for high-precision radial velocity follow-up observations to obtain dynamical masses of any planet companions found by TESS.

\begin{figure*}
\centering \includegraphics[width=0.96\textwidth]{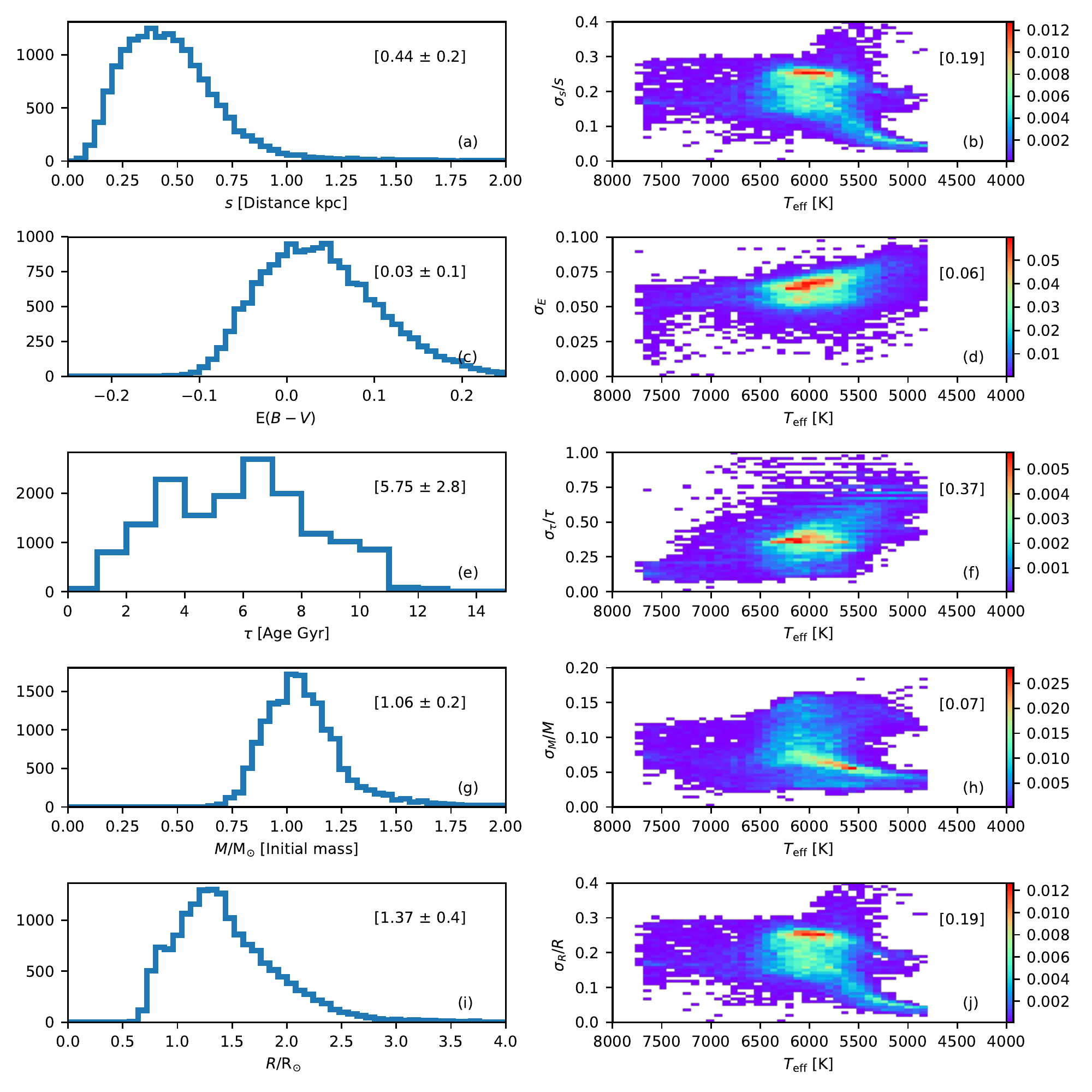}\caption{Stellar properties of the hot dwarf sample derived using spectroscopic parameters and isochrones (BSTEP, \autoref{properties}). Left panels: Distribution of stellar properties. The population median and standard deviation values are shown in brackets.  Right panels: Distribution of uncertainty in stellar properties as a function of temperature.  Color indicates density of data points. The population median uncertainty is shown in brackets.
\label{fig:gbm_summary}}
\end{figure*}
The distribution of stellar properties (distance, extinction, age, initial-mass, and radius) and their uncertainties, derived using our BSTEP isochrones-based approach is shown in \autoref{fig:gbm_summary}.  We note that the uncertainties are internal and do not include any systematics related to for exmaple the adopted stellar model grid.  From \autoref{fig:gbm_summary}a, we see that the stars in our sample are relatively nearby, with a median distance of 450 pc.  This is expected because the stars are dwarfs and the TESS-HERMES survey is limited to relatively bright stars.  Typical uncertainties on distance is about 19\%, but shows a strong dependence on temperature and is significantly lower for cool dwarfs (\autoref{fig:gbm_summary}b).  The median extinction is quite low, but has a large spread, so a good fraction of stars have extinction greater than 0.1 (\autoref{fig:gbm_summary}c).  $E(B-V)$ estimates have an uncertainty of 0.06 with almost no dependence on temperature (\autoref{fig:gbm_summary}d).  Stars span in age from 0 to 13 Gyr, but the age distribution peaks at around 6 Gyr (\autoref{fig:gbm_summary}e). The age estimates have an uncertainty of about 37\% (\autoref{fig:gbm_summary}f).  Most stars have a mass close to solar (\autoref{fig:gbm_summary}g).  The mass estimates are also quite precise (typical internal uncertainty of 7\%) (\autoref{fig:gbm_summary}h).  The median radius of the stars is 1.38\rsol, but the distribution is quite asymmetric with a long tail extending to higher values due to the subgiants (\autoref{fig:gbm_summary}i).  The uncertainty on radius is about 19\% and shows a strong dependence on temperature (\autoref{fig:gbm_summary}j) just like the uncertainty on distance (\autoref{fig:gbm_summary}a).  In fact, the uncertainty on radius and distance are very similar to each other. This is because the radius and distance of a star are both related to its luminosity for a fixed temperature.

\subsection{Benchmarking spectroscopic distances with trigonometric parallaxes}

\begin{figure}
\centering \includegraphics[width=0.48\textwidth]{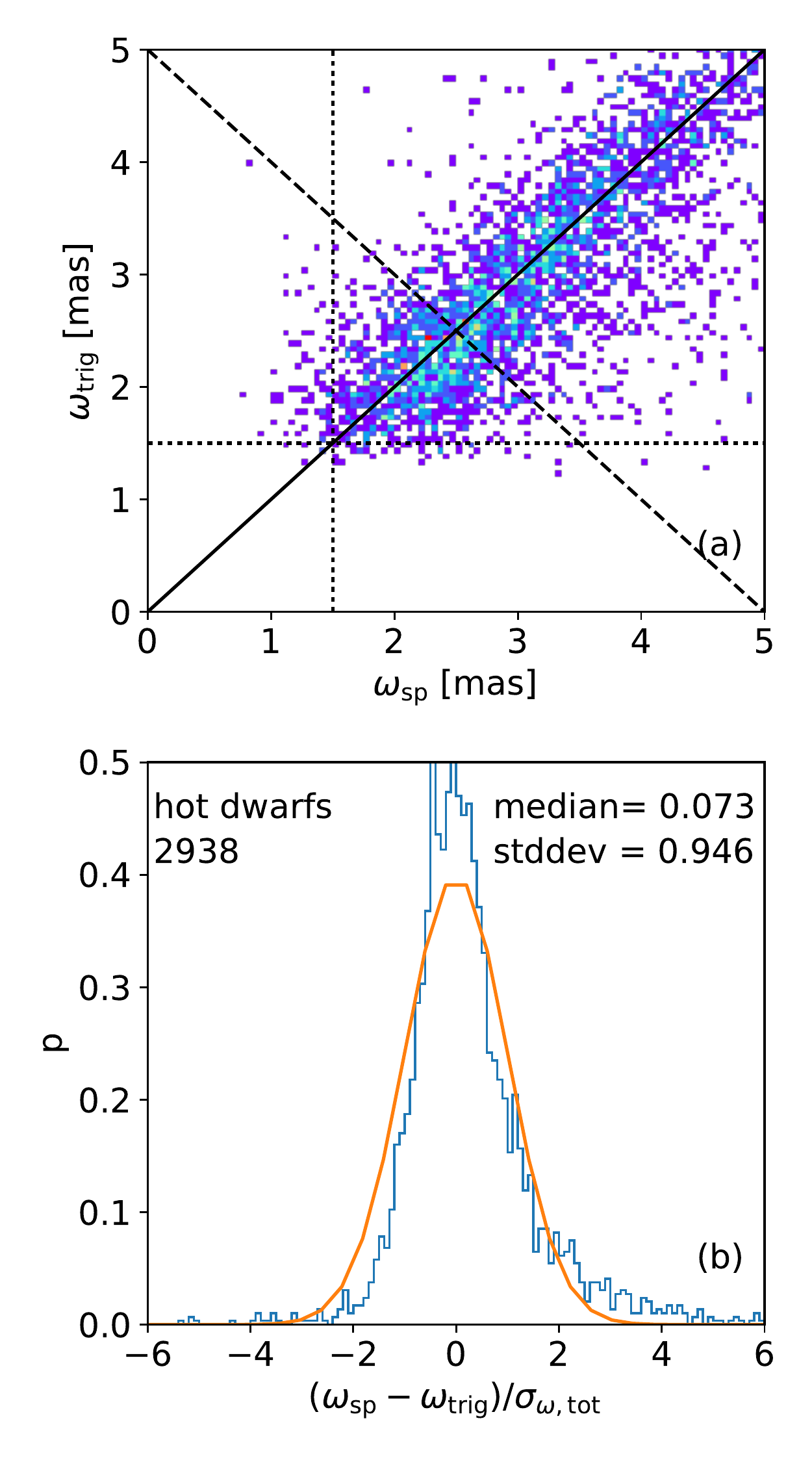}\caption{Comparison of inferred spectroscopic parallax, $\omega_{\rm sp}$, with the trigonometric parallax, $\omega_{\rm trig}$, from Hipparcos or Gaia for our dwarf/subgiant sample.  (a) $\omega_{\rm sp}$ versus $\omega_{\rm trig}$ where color indicates density of data points.  The solid diagonal line shows the 1-to-1 relation.  The vertical and horizontal dotted lines indicate the cuts in parallax uncertainty of 20\%, and the diagonal dashed line shows the third cut to avoid biasing the comparison shown in panel (b).  (b) Distribution of the difference between $\omega_{\rm sp}$ and $\omega_{\rm trig}$ in units of the uncertainty in the difference (blue curve).  A normal distribution with zero mean and standard deviation of unity is shown for reference (orange curve).
\label{fig:omega_sp_tgas}}
\end{figure}

In this section, we use trigonometric parallaxes, $\omega_{\rm trig}$, from the Hipparcos and Gaia DR1 catalogs (as given in the TIC) to test the accuracy of our spectroscopic-based parallaxes, $\omega_{\rm sp}$.  For each star we simply take the trigonometric parallax adopted by the TIC.  We restrict our sample to stars with the most precise parallaxes, specifically $\sigma_{\omega,{\rm trig}}/\omega> 0.2$, which roughly translates into those with $\omega_{\rm trig} > 5 \langle \sigma_{\omega,{\rm trig}}\rangle$, providing a set of approximately 3700 stars.
Such precision-based selection can potentially introduce biases when comparing $\omega_{\rm sp}$ with $\omega_{\rm trig}$ due to truncation of the sample in $\omega_{\rm sp}$ versus $\omega_{\rm trig}$ space (see \autoref{fig:omega_sp_tgas}a).
To restore symmetry for the comparison, we therefore also restrict the sample with respect to the spectroscopic parallax, $\omega_{\rm sp} > 5 \langle\sigma_{\omega,{\rm trig}}\rangle$. We additionally make a diagonal cut to further limit boundary effects towards low parallax values (\autoref{fig:omega_sp_tgas}a, dashed line).  This cut corresponds to
\be
\frac{\omega_{\rm sp}+\omega_{\rm trig}}{2}-2\langle\sigma_{\omega,{\rm tot}}\rangle & > &  5 \langle\sigma_{\omega,{\rm trig}}\rangle,
\ee
or after rearranging, $\omega_{\rm sp}> 5 -\omega_{\rm trig}$, (using $\langle\sigma_{\omega,{\rm trig}}\rangle \sim 0.3$ and $\langle\sigma_{\omega,{\rm tot}}\rangle \sim 0.5$)

We now plot, in \autoref{fig:omega_sp_tgas}b (blue curve), the distribution of $(\omega_{\rm sp}-\omega_{\rm trig})/\sigma_{\omega,{\rm tot}}$, where $\sigma_{\omega,{\rm tot}}=\sqrt{\sigma^2_{\omega,{\rm sp}}+\sigma^2_{\omega,{\rm trig}}}$, for the stars that fulfill the cuts illustrated by the dotted and dashed lines in \autoref{fig:omega_sp_tgas}a.
This should be a normal distribution with zero mean and unit standard deviation if both estimates of parallax are in agreement with each other (orange curve).  We find that the bias is small, less than 10\% of the uncertainty, which validates our spectroscopic pipeline and the BSTEP pipeline for estimating stellar properties.  Nevertheless, the bias is statistically significant, at the 4$\sigma$-level, and suggests that there is still room for improvement.

In Table~\ref{tab:catalog} we describe the content of each column in the TESS-HERMES catalog.  It contains a re-derived priority based on the priority in the TIC, but corrected using the equation given in Section 3.3 of \citet{2017arXiv170600495S} and adopting the radii that we derive in this work. The catalog can be accessed via \href{http://www.physics.usyd.edu.au/tess-hermes/}{http://www.physics.usyd.edu.au/tess-hermes/} , or at MAST (\href{https://archive.stsci.edu/prepds/tess-hermes/}{https://archive.stsci.edu/prepds/tess-hermes/}).
\begin{table*}
\caption{Catalog Description}
\begin{tabular}{l l l l l}
\hline\hline
Name      & Description                   & Units       & Datatype & Source \\
\hline
ra        & Right Ascension (ICRS,Epoch=J2000)       & deg         & float64  & UCAC4 \\
dec       & Declination     (ICRS,Epoch=J2000)       & deg         & float64  & UCAC4 \\
tic\_id   & TESS input catalog identifier            &             & int64    & TIC \\
tmass\_id & 2MASS identifier                         &             & char[16] & UCAC4 \\
ucac      & UCAC4 identifier                         &             & char[10] & 2MASS \\
sobject\_id  & Galah Spectrum identifier             &             & int64    & Galah 1.3, Sect. 2.2\\
snr\_c1   & S/N per pixel for ccd-1 (blue channel)   &             & float64 & Galah 1.3, Sect. 2.2 \\
snr\_c2   & S/N per pixel for ccd-2 (green channel)  &             & float64 & Galah 1.3, Sect. 2.2 \\
snr\_c3   & S/N per pixel for ccd-3 (red channel)    &             & float64 & Galah 1.3, Sect. 2.2 \\
snr\_c4   & S/N per pixel for ccd-4 (infrared channel)&            & float64 & Galah 1.3, Sect. 2.2 \\
jmag      & 2MASS $J$ magnitude           & mag         & float64 & 2MASS \\
hmag      & 2MASS $H$ magnitude           & mag         & float64 & 2MASS \\
kmag      & 2MASS $Ks$ magnitude          & mag         & float64 & 2MASS \\
teff      & Effective temperature  \teff\ & K           & float64 & Galah 1.3, Sect. 2.3 \\
logg      & Surface gravity $\log g$      & dex         & float64 & Galah 1.3, Sect. 2.3 \\
feh       & Iron abundance  [Fe/H]        & dex         & float64 & Galah 1.3, Sect. 2.3 \\
vmic      & Microturbulence $\xi$         & km/s        & float64 & Galah 1.3, Sect. 2.3 \\
vsini     & Rotational and Macroturbulence Velocity $v_{\rm mac+rot}$& km/s        & float64 & Galah 1.3, Sect. 2.3 \\
mini      & Initial mass $m_{\rm ini}$    & M$_{\odot}$ & float64 & Isochrone-based Bayesian estimates, Sect. 3  \\
mact      & Actual mass $m_{\rm act}$ (incl. mass loss)& M$_{\odot}$ & float64 & Isochrone-based Bayesian estimates, Sect. 3  \\
radius    & Stellar radius $R_*$          & R$_{\odot}$ & float64 & Isochrone-based Bayesian estimates, Sect. 3  \\
dist      & Distance $s$                  & kpc         & float64 & Isochrone-based Bayesian estimates, Sect. 3  \\
ebv       & Extinction $E(B-V)$           & mag         & float64 & Isochrone-based Bayesian estimates, Sect. 3  \\
age       & Age $\tau$                    & Gyr         & float64 & Isochrone-based Bayesian estimates, Sect. 3  \\
priority\_tic   & Priority in the TIC/CTL &             & float64 & TIC \\
stflag\_hermes & Stellar type flag        &             & char[9] & One of the following ['lmc', 'giant', 'cooldwarf', 'hotdwarf']\\
spflag\_hermes & Spectroscopic quality flag &           & int64 & One of the following $[0,1,2,3]$. Sec 2.3 \autoref{tab:spflag}  \\
e\_jmag   & Uncertainty in 2MASS $J$ mag  & mag         & float32 & 2MASS\\
e\_hmag   & Uncertainty in 2MASS $H$ mag  & mag         & float32 & 2MASS\\
e\_kmag   & Uncertainty in 2MASS $K$ mag  & mag         & float32 & 2MASS\\
e\_teff   & Uncertainty in  \teff\        & K           & float64 & Galah 1.3, Sect. 2.3 \\
e\_logg   & Uncertainty in $\log g$       & dex         & float64 & Galah 1.3, Sect. 2.3 \\
e\_feh    & Uncertainty in [Fe/H]         & dex         & float64 & Galah 1.3, Sect. 2.3 \\
e\_vmic   & Uncertainty in microturbulence & km/s       & float64 & Galah 1.3, Sect. 2.3 \\
e\_vsini  & Uncertainty in $v\sin i$      & km/s        & float64 & Galah 1.3, Sect. 2.3 \\
e\_mini   & Uncertainty in initial mass   & M$_{\odot}$ & float64 & Isochrone-based Bayesian estimates, Sect. 3  \\
e\_mact   & Uncertainty in actual mass    & M$_{\odot}$ & float64 & Isochrone-based Bayesian estimates, Sect. 3  \\
e\_radius & Uncertainty in stellar radius & R$_{\odot}$ & float64 & Isochrone-based Bayesian estimates, Sect. 3  \\
e\_dist   & Uncertainty in distance       & kpc         & float64 & Isochrone-based Bayesian estimates, Sect. 3  \\
e\_ebv    & Uncertainty in extinction     & mag         & float64 & Isochrone-based Bayesian estimates, Sect. 3  \\
e\_age    & Uncertainty in age            & Gyr         & float64 & Isochrone-based Bayesian estimates, Sect. 3  \\
\hline
\end{tabular}
\label{tab:catalog}
\end{table*}

\section{Summary}\label{summary}

We have presented the first observations of the TESS-HERMES survey aimed at facilitating the scientific output of the TESS mission by high-resolution spectroscopy of up to 40,000 stars.  All our targets range $10<V<13.1$ and fall within a 13 degree radius of the southern ecliptic pole, which encapsulates the TESS CVZ that will be observed first by the mission.  The current data release covers the first 81 1-degree radius subfields, or almost 25,000 stars.

We find that our catalog of spectroscopic results is magnitude complete down to 12.1 mag in the TESS band for stars hotter than $4800\,$K.  In addition to spectroscopic values, we also provide fundamental stellar properties inferred from the Bayesian isochrone-based scheme, BSTEP.  Our results confirm that the TIC generally is performing as expected in terms of separating dwarfs and giants.  However, we do find that a significant fraction of cool dwarfs are flagged as giants in the TIC, suggesting these targets should in fact be given high priority for the search of exoplanets.  In addition, our spectroscopic results enables clear radius differentiation among dwarfs and subgiants.  We advocate that our estimated radii be used to re-prioritize the stars in the CTL to maximize the scientific output from the limited number of 2-minute cadence slots in the CVZ.  As an interesting sideline, we see clear evidence in the spectroscopy of high radial velocity stars in the sample, which we suspect belong to the LMC.

When finished, the TESS-HERMES survey will have observed 130 1-degree radius subfields covering 408 square degrees around the ecliptic pole providing a 77\% sky coverage of the southern CVZ of TESS.  The remaining six nights required to complete the survey coverage outlined in \autoref{fig:tess_fields} have been granted, and observations will be carried out at the end of December 2017, after which we expect to publish the completed TESS-HERMES CVZ catalog.

\acknowledgments
S.S. is funded by University of Sydney
Senior Fellowship made possible by the office of the Deputy Vice Chancellor of Research, and partial funding from Bland-Hawthorn's Laureate Fellowship from the Australian Research Council.
D.S. is the recipient of an Australian Research Council Future Fellowship (project number FT1400147).  S.B. and K.L. acknowledge funds from the Alexander von Humboldt Foundation in the framework of the Sofja Kovalevskaja Award endowed by the Federal Ministry of Education and Research. K.L. acknowledges funds from the Swedish Research Council (Grant nr. 2015-00415\_3) and Marie Sklodowska Curie Actions (Cofund Project INCA 600398).  T.Z. acknowledges financial support from the Slovenian Research Agency (research core funding No. P1-0188).  A.~R.~C. is supported through Australian Research Council Discovery Project grant DP160100637.
Parts of the computations were performed on resources provided by the Swedish National Infrastructure for Computing (SNIC) at UPPMAX under project 2015/1-309 and 2016/1-400.
We thank Keivan Stassun and Willie Torres for helpful comments on the manuscript.
This work has made use of data from the European Space Agency (ESA) mission {\it Gaia} (\url{https://www.cosmos.esa.int/gaia}), processed by the {\it Gaia} Data Processing and Analysis Consortium (DPAC, \url{https://www.cosmos.esa.int/web/gaia/dpac/consortium}). Funding for the DPAC has been provided by national institutions, in particular the institutions participating in the {\it Gaia} Multilateral Agreement.

\facilities{AAT}
\software{Numpy, Matplotlib}

\bibliographystyle{yahapj}

\end{document}